\documentclass[preprint2,twocolumn,times,tighten]{aastex62}

\usepackage{color}
\usepackage{enumitem}
\usepackage{hyperref}
\usepackage{verbatim}
\usepackage{amsmath,amstext,mathrsfs}
\usepackage[all]{hypcap} 
\usepackage{afterpage}    
\usepackage{marginnote}
\usepackage{makecell}
\usepackage{subfigure}

\graphicspath{{./}{figures/}}

\shorttitle{Predicting dust polarization using \ion{H}{1}}
\shortauthors{Hu, Yuen \& Lazarian}

\begin{document}
	
	\title{Predictions of Cosmic Microwave Background Foregrounds Dust Polarization Using Velocity Gradients}
	
	\correspondingauthor{A. Lazarian}
	\email{alazarian@facstaff.wisc.edu}
	
	\author[0000-0002-8455-0805]{Yue Hu}
	\affiliation{Department of Physics, University of Wisconsin-Madison}
	\affiliation{Department of Astronomy, University of Wisconsin-Madison}
	
	\author[0000-0003-1683-9153]{Ka Ho Yuen}
	\affiliation{Department of Astronomy, University of Wisconsin-Madison}
	
	\author{A. Lazarian}
	\affiliation{Department of Astronomy, University of Wisconsin-Madison}

	\begin{abstract}
		The observations of fluctuations in the cosmic microwave background provide information about primordial inhomogeneities in the universe. However, the B-mode polarization of the inflationary gravitational wave is contaminated by the Galactic foreground polarized radiation arising from dust aligned by interstellar magnetic fields. To trace magnetic fields we use the Velocity Gradient Technique, which employs a modern understanding of the nature of magneto-hydrodynamics turbulent motions. In this paper, we combine the VGT with the Principal Component Analysis (PCA) to improve the accuracy of magnetic field tracing. We apply the VGT-PCA to the high-resolution neutral hydrogen data from the GALFA-\ion{H}{1} survey to predict the polarization of dust. We report that the predicted directions of dust polarization provide good correspondence with those reported by Planck 353GHz, the alignment measure between the two $AM \simeq 0.79\pm0.01$. We show that our results statistically agree with the Planck polarization in terms of magnetic field tracing. We find that the variation of dust emission efficiency across the sky is small. Using our maps of predicted polarization we calculate the ratio of the E and B modes and show that  $BB/EE \simeq 0.53\pm0.10$, which is similar to the result from Plank polarization.
	\end{abstract}
	
	
	\section{Introduction} 
	\label{sec:intro}
	The fluctuations in the cosmic microwave background (CMB) polarization, which can be decomposed into "electric" (E) and "magnetic" (B) components and those  contain important information of the evolution of the early universe \citep{2016A&A...586A.141P,2003PhRvD..68h3509L,2017ApJ...846...45M, 2016A&A...594A..11P}. In particular, the measurement of CMB B-modes offers the possibility to study the cosmological origin of inflationary gravitational waves \citep{2016ARA&A..54..227K,2014JCAP...06..053F,2016A&A...586A.141P,2014PhRvL.112x1101B}. However, the polarized thermal emission from diffuse Galactic dust is the main foreground present in measurements of the CMB polarization at frequencies above 100 GHz \citep{2014A&A...571A..11P,2015A&A...576A.105P,2018arXiv180104945P}. The CMB B-mode signal is therefore contaminated by the Galactic dust foreground polarization arising from complicated interstellar magnetic fields \citep{2015PhRvD..91h1303K,2016MNRAS.462.2343V,1989ApJ...346..728J}. To get insight into the CMB B-mode signal, a comprehensive picture of Galactic polarized foreground is essential to disentangle the spectral energy distribution of dust and CMB polarization across frequencies.
	The recent Plank survey gives full-sky polarization maps of dust emission \citep{2018arXiv180104945P}. However, the polarization fraction is theoretically determined by the dust column along the line of sight and the angle of the mean magnetic field concerning the plane of the sky \citep{2014A&A...571A..11P}. At high Galactic latitude regions, dust emissivity is low \citep{2014A&A...566A..55P}, and polarization fraction is minimal \citep{2016A&A...594A..10P}. It is, therefore, challenging to characterize the complete Galactic foreground polarization using polarized dust emission. Nevertheless, even this minimal galactic polarization can present a challenge in measuring the weak B mode signal. 
	
	The approach of studying magnetic fields with the Velocity Gradients Technique (VGT), has been a fast-developing branch of interstellar medium magnetic field studies. For instance, the Velocity Centroid Gradients (VCGs, \citealt{2017ApJ...835...41G, 2017ApJ...837L..24Y,2017arXiv170303026Y}) was first proposed to trace the magnetic field using the fact that in magnetohydrodynamic (MHD) turbulence eddies are elongated along with magnetic field direction surrounding the eddis \citep{1995ApJ...438..763G,1999ApJ...517..700L}. Aside from the velocity centroids, the use of the velocity information contained in the thin velocity channel map is also explored by \cite{LP00} and \cite{2019arXiv190403173Y}. The velocity gradients in such narrow velocity channels are shown to be more accurate in tracing magnetic fields. With that development, the VGT has already been successfully applied to a wide range of column densities from diffuse transparent gas \citep{2017ApJ...837L..24Y, PCA,2019ApJ...874...25G,IGs} to molecular self-absorbing dense gas \citep{survey,velac} as well as the estimation of magnetization level \citep{2018ApJ...865...46L,survey}. Although the practical application of VGT is affected by the quality of data \citep{2018ApJ...865...54Y}, \citet{PCA} explored the noise suppression method for VGT and showed that the Principal Component Analysis (PCA) could increase the accuracy of VGT in tracing magnetic fields. 
	
	The PCA is originally a technique used in image processing and image compression. Regarding astrophysical applications, the PCA was applied to obtain the turbulence spectrum from observations \citep{2002ApJ...566..276B,2002ApJ...566..289B}, as well as studying turbulence anisotropies \citep{2008ApJ...680..420H}. Since the principal components of gradients decomposed by PCA are more anisotropic, we use the gradients of the decomposed principal components as an alternative tool to trace the interstellar magnetic fields and predict the Galactic foreground polarization, denoted as VGT-PCA technique. 
	
	In what follows, we illustrate the theoretical foundation of tracing magnetic field based on MHD turbulence anisotropy, in \S\ref{sec:theory}. In \S\ref{sec:method}, we describe the full algorithm in implementing the VGT-PCA. In \S\ref{sec:results}, we apply VGT-PCA to a vast range of  GALFA-\ion{H}{1} data and make comparisons with Planck polarization. In \S\ref{sec.dis}, we discuss the modeling of the three-dimensional Galactic magnetic field, the possible application, and further development of VGT-PCA. In \S\ref{sec:con}, we give our conclusions.
	
	\section{Theoretical Consideration}
	\label{sec:theory}
	\subsection{Theory of MHD turbulence}
	\citet{1995ApJ...438..763G} established the theoretical foundation of magnetic fields tracing techniques through MHD turbulence statistics. Considering the Kolmogorov cascade with the injection velocity $v_l\simeq l_\perp^{\frac{1}{3}}$, \cite{1995ApJ...438..763G} predicted that incompressible MHD turbulence is anisotropic, i.e. turbulent eddies are elongated along the direction of magnetic field. The derivation in \cite{1995ApJ...438..763G} is provided in the frame of the mean field, which was also the accepted frame for all theoretical constructions prior to the study.
	
	However, the theory of turbulent reconnection in \citet{1999ApJ...517..700L} showed that the GS95 relations valid not in the mean field reference frame, but in the reference frame of individual eddies. In \citet{1999ApJ...517..700L}, the fast turbulent reconnection theory explained that magnetic tension force resists any other types of magnetic field motion, except the rotating motion of eddies together with magnetic field lines in the direction perpendicular to magnetic fields {\it local} to the eddy. For the VGT the concept of local magnetic field frame is crutial. This concept was supported by numerical simulations and is well established by now  \citep{2000ApJ...539..273C,Cho2002,Cho2003CompressibleImplications}. 
	
	It is because the turbulent motions are aligned with the local direction of magnetic field, that studying velocity gradients one can obtain the detailed structure of magnetic field, not just its mean direction. Understanding $\bot$ and $\|$ in this local direction sense, one can use the GS95 relation between the parallel and perpendicular eddy size, i.e.
	$l_{||}\simeq l_{\perp}^{\frac{2}{3}}$, to see that the motions within the smallest eddies are the most aligned with the local magnetic fields.  
	
	Considering the eddies perpendicular that due to turbulent reconnection freely mix magnetic field perpendicular to its direction, one can easily understand that the velocity fluctuations are expected to follow the the Kolmogorov scaling $V_{l,\bot} \sim l_{\bot}^{1/3}$ and therefore the gradients are going to be the largest at the smallest scale $v_{l,\perp}/l_{\perp}\simeq l_{\perp}^{-2/3}$.
	It is also obvious that the measured velocity gradients are perpendicular to the magnetic field at the smallest resolved scales, i.e., they well trace the magnetic field in the turbulent volume. Note, that similar to the case of far-infrared polarimetry, one should turn the direction of gradients by 90$^\circ$ to obtain magnetic field direction.
	
	The statistics of velocity fluctuation is not directly available from observations. To get insight into velocity statistics, \citet{LP00,LP04,2017MNRAS.464.3617K} developed the theory of statistics of the Position-Position-Velocity (PPV) spectroscopic data cubes. They showed that intensity fluctuations in PPV cubes can arise due to turbulent velocities along the line of sight. This effect of velocity crowding causes velocity caustics. It was shown that when velocity channel width $\Delta v$ and velocity dispersion $\delta v$ satisfy the criterion:
	\begin{equation}
		\label{eq1}
		\Delta v^2<\delta v^2
	\end{equation}
	this velocity channel is considered a thin velocity channel. Velocity fluctuation is can dominate in this thin channel due to the effect of velocity caustics. Based on this theory, \cite{2018ApJ...865...46L} and \cite{IGs} proposed to use fluctuations of velocity within thin velocity channel maps to reveal the information of velocity gradients.
	
	\begin{figure*}
		\centering
		\includegraphics[width=.99\linewidth,height=0.45\linewidth]{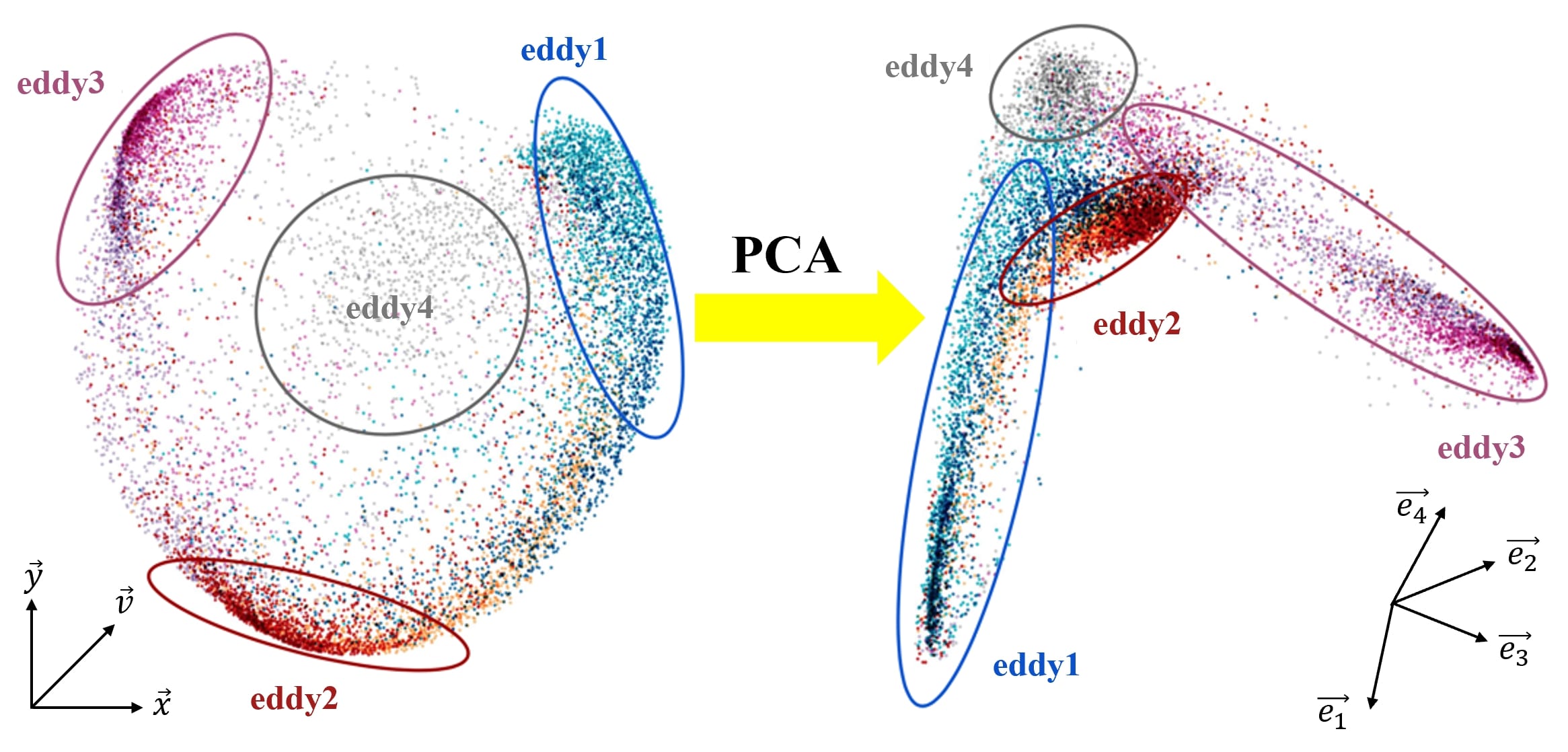}
		\caption{\label{fig:pca} An illustration of how PCA works on PPV cubes. Supposing there exist four crucial velocity components, i.e. eddy1/2/3/4, in original PPV cube (i.e., in the $\vec x$-$\vec y$-$\vec v$ space, see the left panel), the implementation of PCA converts these components to a set of linearly independent variables called principal components (see the right panel). The principal components locate at a new orthogonal basis formed by their corresponding eigenvectors (i.e., the $\vec e_1$-$\vec e_2$-$\vec e_3$-$\vec e_4$ space), which are oriented along the direction of the large axis of each principal component respectively. The length of the large axis is positively proportional to its corresponding eigenvalue, whose physical meaning is closely related to the value of the turbulence velocity dispersion $\delta v^2$. In particular, those larger eddies correspond to the largest-scale contributions of turbulence eddies along the line of sight $\delta v^2 \simeq l^{\frac{2}{3}}$, where $l$ is the scale of the eddy. The significance of eddy1/2/3/4 in the new space is therefore enhanced through PCA.}
	\end{figure*}
	\subsection{Polarized thermal dust emission}
	Measuring the fraction of polarized thermal dust emission deepens our understanding of the diffuse ISM and also helps for the search for inflationary gravitational wave B-mode polarization in the CMB. The Stokes parameters I, Q, U are describing the polarization state of thermal dust emission. For constant magnetic field orientation and polarization fraction along the line-of-sight (LOS), the integral equations of the Stokes parameters for linear dust polarization are defined as:
	\begin{equation}
		\label{eq.1}
		\begin{aligned}
			&Q=\int p_{max}Rcos(2\psi)cos^2\gamma dI\\
			&U=\int p_{max}Rsin(2\psi)cos^2\gamma dI\\
			&p=\frac{\sqrt{Q^2+U^2}}{I}=p_{max}RFcos^2\gamma
		\end{aligned}
	\end{equation}
	where $\gamma$ is the angle between magnetic field and the plane-of-the-sky (POS), $\psi$ is polarization angle, $p_{max}$ is the maximum value of polarization fraction, R is the Rayleigh reduction factor (i.e. the efficiency of grain alignment), and F is the depolarization factor \citep{2015A&A...576A.105P}. The theory of grain alignment predicts (see Lazarian \& Hoang 2007) that the radiative torques (RATs) are capable of well aligning grains with magnetic field in diffuse interstellar medium, provided that the grains precess fast in the ambient magnetic field. This is the case of silicate grains, but not for the silicate ones (see Lazarian \& Hoang 2019).  
	
	Once the Stokes parameters, $Q(\hat{n})$ and $U(\hat{n})$, have been measured as functions of spherical polar coordinates $\theta$, $\phi$ (where $\hat{n}$ denotes a unit vector pointing at polar angle $\theta \in [0, \pi]$ and azimuth $\phi \in [0, 2\pi]$) on the full sky polarization, we have the polarization tensor:
	\begin{equation}
		P(\hat{n})=\frac{1}{\sqrt{2}}
		\begin{bmatrix}
			Q(\hat{n}) & U(\hat{n})sin\theta  \\
			U(\hat{n})sin\theta & -Q(\hat{n})sin^2\theta \\
		\end{bmatrix}
	\end{equation}
	The polarization tensor field on the sphere can be expanded in terms of basis functions that are
	gradients and curls of spherical harmonics $Y_{lm}(\hat{n})$:
	\begin{equation}
		P(\hat{n})=\sum^{\infty}_{l=2} \sum^{l}_{m=-l}[a_{lm}^{E}Y_{lm}^{E}(\hat{n})+a_{lm}^{B}Y_{lm}^{B}(\hat{n})]
	\end{equation}
	The expansion coefficients are given by:
	\begin{equation}
		\begin{aligned}
			a_{lm}^{E}=\int P(\hat{n})Y_{lm}^{E*}(\hat{n})d\hat{n}\\
			a_{lm}^{B}=\int P(\hat{n})Y_{lm}^{B*}(\hat{n})d\hat{n}\\
		\end{aligned}
	\end{equation}
	The angular cross power spectra are now:
	\begin{equation}
		\langle a_{lm}^{X*}a_{l'm'}^{X'}\rangle=C_{l}^{XX'}\delta_{ll'}\delta_{mm'}
	\end{equation}
	where X = \{E; B\}, $\langle...\rangle$ denotes the ensemble average, and $\delta$ is the delta function. In general, both the foreground dust polarization and CMB polarization contribute to the spectra. However, it is difficult to separate the contribution from CMB E- and B-mode as the expected signals from inflation and late-time reionization are expected to be small. In this case, accurate assessment and modeling of the Galactic foreground are crucial to the subtraction of the foreground dust polarization to $C_l^{EE}$ and $C_l^{BB}$. 
	
	\section{Methodology}
	\label{sec:method}
	\subsection{Principal Gradient Component Analysis}
	The Principal Component Analysis (PCA) is widely used in image processing and image compression, which effectively decomposes an image of size $N^2$ into $n<N$ eigenmaps. PCA uses an orthogonal linear transformation to convert a set of possibly correlated variables to a set of linearly independent variables called principal components \citep{Hotel33}. Intuitively, PCA can be thought of as fitting a n-dimensional ellipsoid to the data, where each orthogonal axis of the ellipsoid represents a principal component. This fitting in PPV cube is implemented through the calculation of the covariance matrix, which giving the covariance between each velocity channel. The eigenvalues of the covariance matrix correspond to the length of the large axis of each principal component and the eigenvectors are oriented along the direction of the large axis. If some eigenvalue is small, then the variance along that axis and contribution from its corresponding principal component are also small. We can therefore: (i) remove the noise by omitting small eigenvalues and their corresponding principal components from our representation of the dataset \citep{PCA}; (ii) enhance the contribution from crucial components by projecting the original dataset into the new orthogonal basis formed by the eigenvectors. 
	
	As illustrated in Fig.~\ref{fig:pca}, we suppose there exit four kinds of balls with different colors blue, red, purple, and grey mixing in a pool. In PCA's picture, the ball represents each data point in the original PPV cube and the four colors correspond to four crucial velocity components eddy1/2/3/4. The question is how to classify those balls based on their colors (or how to separate those eddies through PCA). To solve it in our daily life, the simplest way is to prepare four boxes firstly and label them by colors (blue, red, purple, or grey) respectively. We randomly pick up a ball and distinguish its color, which equivalents to the calculation of the covariance matrix. We then drop each individual ball into the corresponding box based on its color (e.g., the dropping step equivalents the projection of the original dataset into the new orthogonal basis). After this classification, we find that we need much larger boxes (i.e., the size of the box represents the corresponding eigenvalue) to store blue or purple balls than red or grey balls. We can then say that the number of blue and purple balls are much larger than the number of red and grey balls in the pool, i.e., the contribution from eddy1 and eddy3 is, therefore, more significant than the one from  eddy2 and eddy4, and there is little change in the pool even if we remove red and grey balls. More importantly, in these new classified boxes, we can play with single-color balls without the affect from the others.
	
	In the world of PCA, we project those eddies into a new orthogonal egien-space, i.e., the $\vec e_1$-$\vec e_2$-$\vec e_3$-$\vec e_4$ space, by weighting the original PPV cube with the eigenvectors calculated through PCA. In the new space, the length of the eddy's large axis is positively proportional to its corresponding eigenvalue, whose physical meaning is closely related to the value of the turbulence velocity dispersion $\delta v^2$. In particular, those larger eddies correspond to the largest-scale contributions of turbulence eddies along the line of sight $\delta v^2 \simeq l^{\frac{2}{3}}$, where $l$ is the scale of the eddy. The significance of eddy1 and eddy3 in the new space is therefore separated and signified through PCA. \citet{PCA} used the PCA as a tool to provide the preliminary processing of the spectroscopic data and they found that the structure in the new egien-space becomes more anisotropic. As velocity gradients scale as $v_{l,\perp}/l_{\perp}\simeq l_{\perp}^{-2/3}$ (see Sec.~\ref{sec:theory}), the highly anisotropic components significantly contribute to the accuracy of velocity gradients in tracing magnetic fields. We, therefore, see the possibility of improving VGT, by combining with PCA which that can extract the most crucial velocity components.
	
	In this work, we follow the assumption used in \cite{PCA}: Treat the PPV cube $\rho(x,y,v)$ as the probability density function of three random variables $x,y,v$, then we define the modified covariance matrix \citep{2002ApJ...566..276B,2002ApJ...566..289B} and the eigenvalue equation for this covariance matrix as:
	\begin{equation}
		\begin{aligned}
			S(v_i,v_j) \propto &\int dxdy \rho(x,y,v_i)\rho(x,y,v_j)\\
			&-\int dxdy \rho(x,y,v_i)\int dxdy\rho(x,y,v_j)
		\end{aligned}
	\end{equation}
	\begin{equation}
		\textbf{S}\cdot \textbf{u}=\lambda\textbf{u}
	\end{equation}
	where \textbf{S} is the co-variance matrix with matrix element $S(v_i,v_j)$, with $i,j=1,2,...,n_v$. $\lambda$ is the eigenvalues associated with the eigenvector matrix $\textbf{u}$. Note that $\textbf{u}$ consists of eigenvectors $\textbf{u}_i$ sorted in a rank based on the decreasing order of $\lambda_i$, with elements $u_{ij}$, $i,j=1,2,...,n_v$. Thereafter, we project the PPV cube into the direction of each eigenvector $\textbf{u}_i$, by weighting channel $\rho(x,y,v_j)$ with the corresponding eigenvector $u_{ij}$. Hence, we get the corresponding eigen-map $I_i(x,y)$:
	
	\begin{equation}
		I_i(x,y)=\sum_j^{n_v}u_{ij}\cdot \rho(x,y,v_j)
	\end{equation}
	
	By repeating the procedure for each eigenvector $\textbf{u}_i$, we finally get a set of eigen-maps $I_i(x,y)$, with $i=1,2,...,n_v$. From an individual eigen-map $I_i(x,y)$, the gradient orientation at individual pixel $(x_{m},y_{n})$ is calculated by convolving the image with 3 $\times$ 3 Sobel kernels:
	$$
	G_x=\begin{pmatrix} 
	-1 & 0 & +1 \\
	-2 & 0 & +2 \\
	-1 & 0 & +1
	\end{pmatrix},
	G_y=\begin{pmatrix} 
	-1 & -2 & -1 \\
	0 & 0 & 0 \\
	+1 & +2 & +1
	\end{pmatrix}
	$$
	\begin{equation}
		\begin{aligned}
			\bigtriangledown_x I_i(x,y)=G_x * I_i(x,y)  \\  
			\bigtriangledown_y I_i(x,y)=G_y * I_i(x,y)  \\
			\psi_{gi}={\rm tan^{-1}}(\frac{\bigtriangledown_y I_i}{\bigtriangledown_x I_i})
		\end{aligned}
	\end{equation}
	where $\bigtriangledown_x I_i(x,y)$ and $\bigtriangledown_y I_i(x,y)$ are the x and y components of gradients respectively. $\psi_{gi}$ is the pixelized gradient map for each $I_i(x,y)$.
	The sub-blocking method, proposed by \citet{2017ApJ...837L..24Y}, is applied after the pixelized gradient map is established. Within a sub-block of interests, the distributions of the gradient vector orientation appear as a more accurate Gaussian profile with the increment of the size of the sub-region. In this case, the mean gradient direction becomes more well-defined, and the alignment between the gradient and magnetic field becomes more accurate. By taking the Gaussian fitting peak value of the gradient distribution in a selected sub-block, we obtain the mean direction of the magnetic field in that sub-region. The sub-block averaging method therefore can increase the significance of important statistical measure and suppresses noise in a sub-region of the gradient field.
	
	In observations, the magnetic field orientation is inferred from the polarization angle $\psi$, which can be derived from the Stokes Q, U maps using the relations:
	
	\begin{equation}
		\begin{aligned}
			&\psi=\frac{1}{2}{\rm tan^{-1}}(\frac{U}{Q})\\
		\end{aligned}
	\end{equation}

	However, the orientation of magnetic field measured by gradients is different from that of dust polarization: gradients measure the sectional magnetic field along the line of sight while that of polarization measures the density-weighted accumulated magnetic field directions along the line of sight. As a result, it is suggestive to compare the Stokes parameter equivalent  Q$_g$ and U$_g$ of the gradient-induced magnetic field to that of polarization:
	\begin{equation}
		\label{eq.11}
		\begin{aligned}
			&Q_g(x,y)=\sum_{i=1}^n I_i(x,y)\cos(2\psi_{gi}(x,y))\\
			&U_g(x,y)=\sum_{i=1}^n I_i(x,y)\sin(2\psi_{gi}(x,y))\\
			&\psi_g=\frac{1}{2}tan^{-1}(\frac{U_g}{Q_g})
		\end{aligned}
	\end{equation}
	where $I_i(x,y)$ is the i-th eigen-map, $\psi_{gi}(x,y)$ is the gradient angle calculated from $I_i(x,y)$ with the sub-block averaging implemented. The  mock polarization angle $\psi_g$ is then defined correspondingly, which gives a probe of plane-of-the-sky magnetic field orientation after rotating 90$^\circ$. Note that in constructing the $Q_g(x,y)$ and $U_g(x,y)$, one can project the data onto the subset of the dominant principal components but not onto all of them, i.e., $n<n_v$, especially when the smallest eigenvalues of the covariance matrix are dominated by noise. In this work, we do not distinguish the noise subset and use $n=n_v$.
	
	\subsection{Alignment Measure}
	We make comparisons with the Planck 353GHz polarized dust signal data from the Planck 3rd Public Data Release (DR3) 2018 of High Frequency Instrument \citep{2018arXiv180706207P}\footnote{Based on observations obtained with Planck (http://www.esa.int/Planck), an ESA science mission with instruments and contributions directly funded by ESA Member States, NASA, and Canada.},  where the signal-to-noise ratio (S/N) of the dust emission is maximum at low Galactic latitude regions. The Planck observations provide Stokes parameter maps I, Q, and U, so the plane-of-the-sky magnetic field orientation angle $\theta$ can be derived from the Stokes parameters: $\theta=\psi-\pi/2$.
	
	The relative orientation between the POS magnetic field predicted by VGT-PCA and the one inferred from Planck polarization is quantified by the \textbf{Alignment Measure (AM)}: 
	
	\begin{align}
		AM=2(\langle {\rm cos^{2}} \theta_{r}\rangle-\frac{1}{2})
	\end{align}
	
	where $\theta_r$ is the angular difference between magnetic field vectors predicted from VGT-PCA and the vector inferred from polarization in a single sub-block. Since AM is insensitive to the sign of the gradient, e.g.~AM($\theta$)\,=\,AM($-\theta$), it is advantageous to test the performance of each method. We expect to get AM = 1 in most cases, which implies a perfect alignment, i.e the rotated $\psi_g$ is parallel to the POS magnetic field. The AM has been widely used in gradients studies \citep{2017ApJ...837L..24Y,survey,velac,PCA,2019ApJ...874...25G,IGs,2017ApJ...835...41G,2017arXiv170303026Y}. Obviously, it is equivalent to $\langle cos2 \theta_{r}\rangle$ that was used late in \citet{CH19}.
	
	\begin{figure*}
		\centering
		\includegraphics[width=.99\linewidth,height=0.49\linewidth]{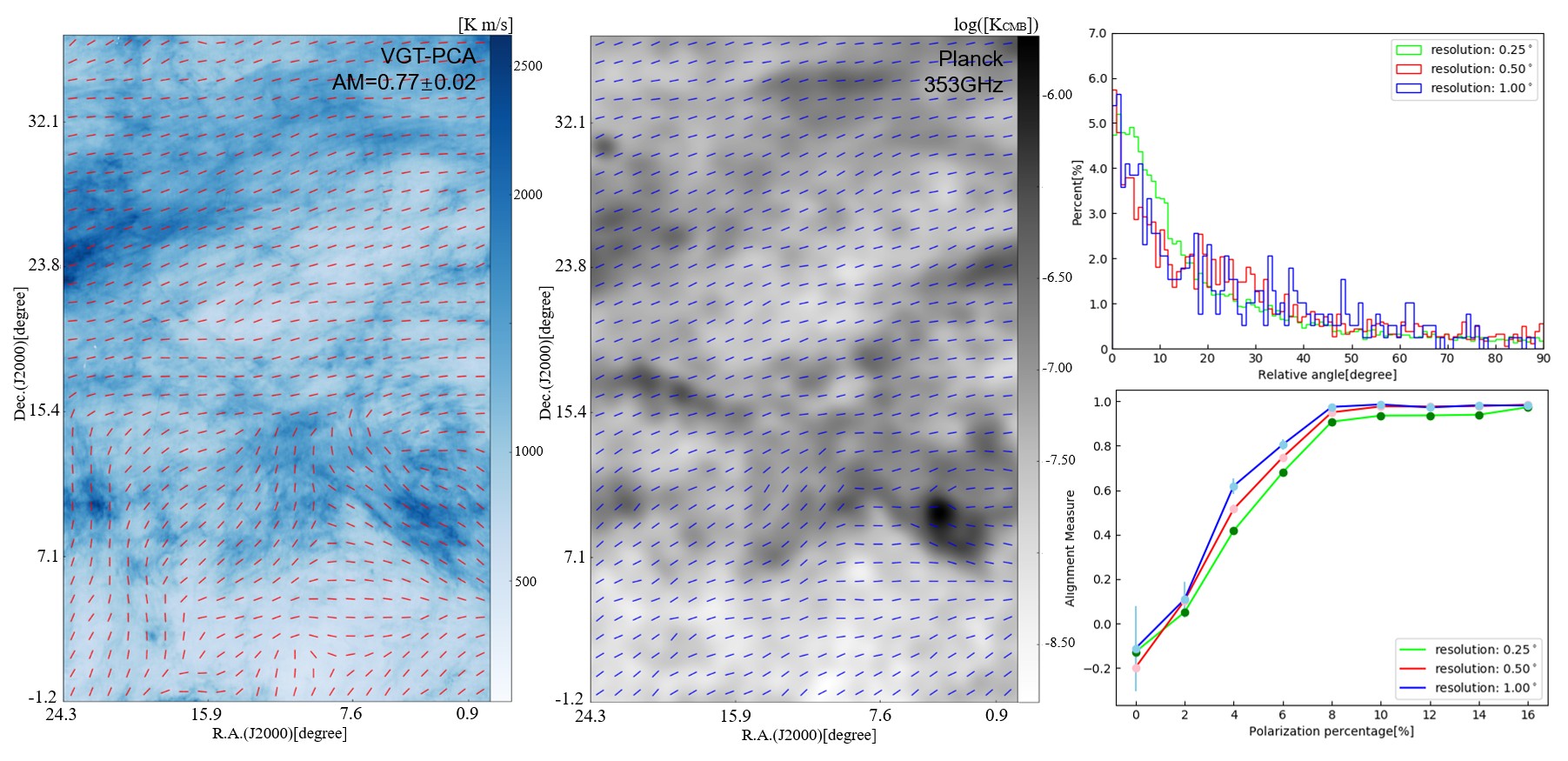}
		\caption{\label{fig:A1} The Morphology of the POS magnetic fields on the sky patch with R.A. from 0.0$^\circ$ to 24.0$^\circ$ and DEC. from -1.2$^\circ$ to 37.1$^\circ$. \textbf{Left:} the magnetic field predicted from VGT-PCA (red segments) with resolution $\simeq 1 ^\circ$. \textbf{Middle:} the magnetic fields inferred from Planck polarization (blue segments). The background color map of VGT-PCA is the GALFA-\ion{H}{1} data integrated intensity from -32 $km\cdot s^{-1}$ to 23 $km\cdot s^{-1}$. \textbf{Right top:} the histogram of the relative orientation between the magnetic field predicted by VGT-PCA and the one inferred from Planck polarization. \textbf{Right bottom:} the variation of the AM with respect to the polarization percentage.}
	\end{figure*}
	
	\begin{figure*}
		\centering
		\includegraphics[width=.99\linewidth,height=0.49\linewidth]{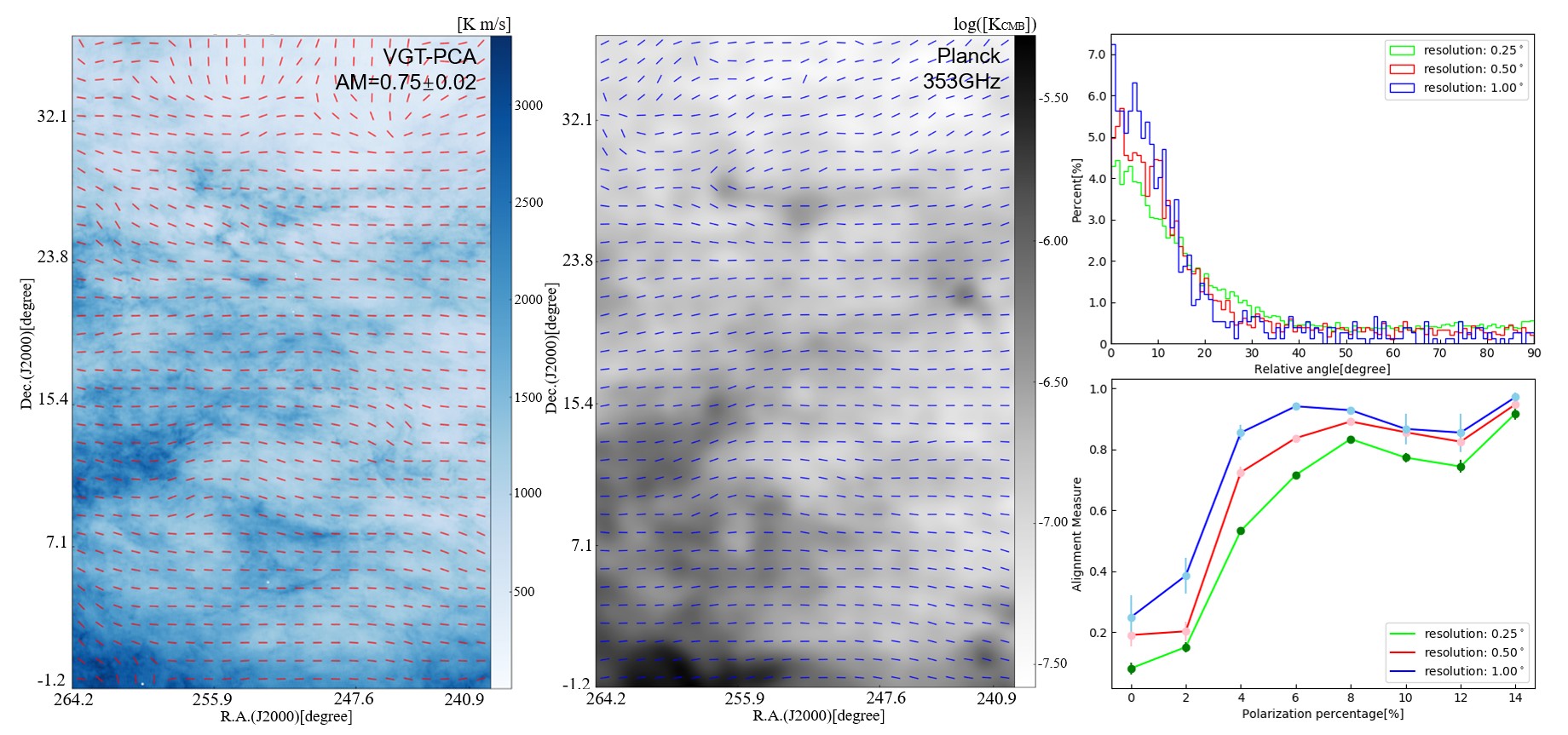}
		\caption{\label{fig:C1} The Morphology of the POS magnetic fields on the sky patch with R.A. from 240.0$^\circ$ to 264.0$^\circ$ and DEC. from -1.2$^\circ$ to 37.1$^\circ$. \textbf{Left:} the magnetic field predicted from VGT-PCA (red segments) with resolution $\simeq 1 ^\circ$. \textbf{Middle:} the magnetic fields inferred from Planck polarization (blue segments). The background color map of VGT-PCA is the GALFA-\ion{H}{1} data intesntiy integrated  from -32 $km\cdot s^{-1}$ to 23 $km\cdot s^{-1}$. \textbf{Right top:} the histogram of the relative orientation between the magnetic field predicted by VGT-PCA and the one inferred from Planck polarization. \textbf{Right bottom:} the variation of the AM with respect to the polarization percentage.}
	\end{figure*}
	
	\begin{figure*}
		\centering
		\includegraphics[width=.99\linewidth,height=1.25\linewidth]{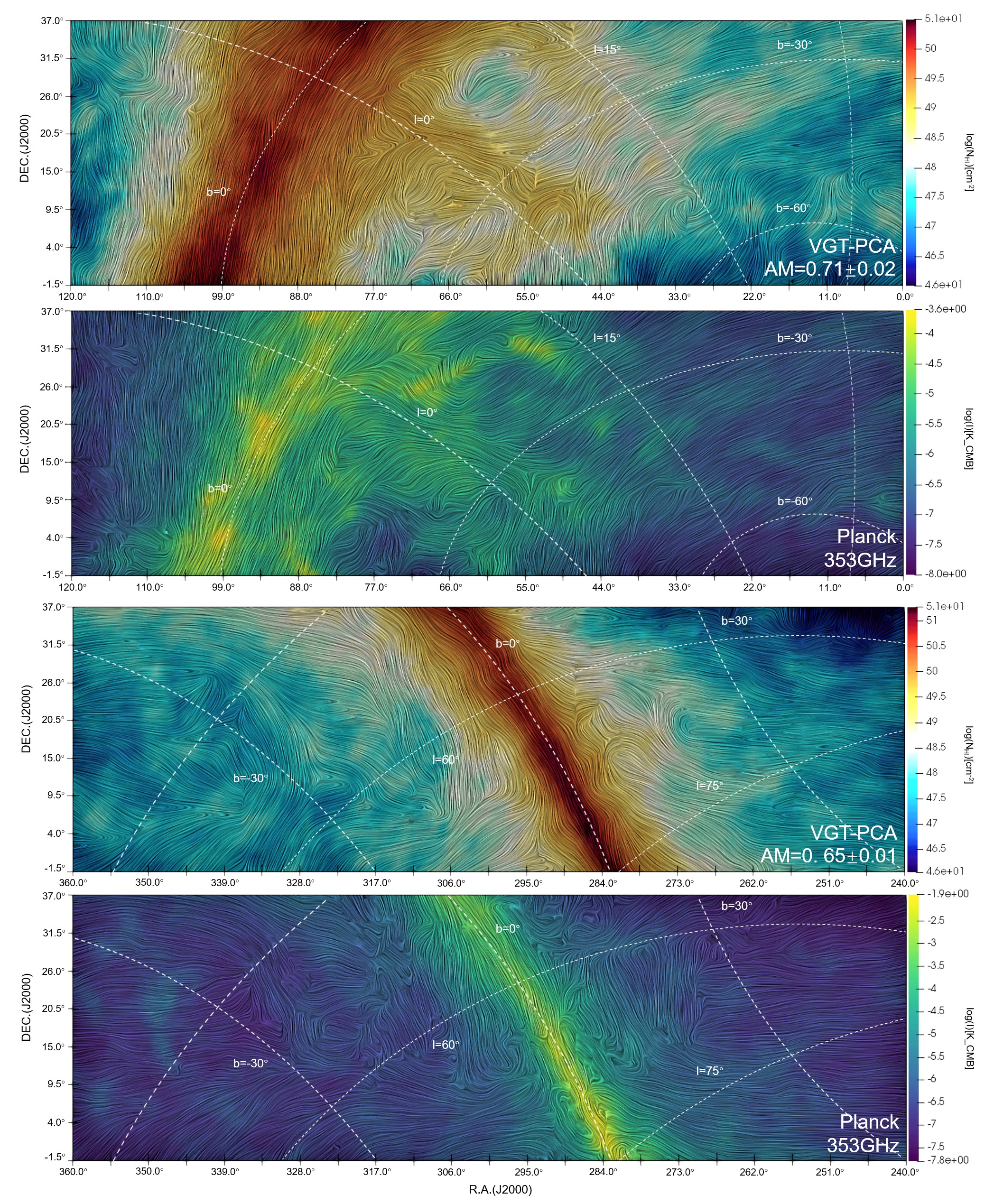}
		\centering
		\caption{\label{fig:PGCA_A} The morphology of the POS magnetic fields at low Galactic latitude regions inferred from Planck polarization (the 2$^{nd}$ and the 4$^{th}$ columns), and VGT-PCA (the 1$^{st}$ and the 3$^{rd}$ columns). The background color maps are of VGT-PCA is the \ion{H}{1} column density map integrated from -90 $km\cdot s^{-1}$ to 90$km\cdot s^{-1}$. The magnetic field is visualized using the Line Integral Convolution (LIC).}
	\end{figure*}
	\section{Results}
	\label{sec:results}
	\subsection{Magnetic fields morphology on low Galactic latitude region}
	\label{sec:low latitude}
	In this work, we use the high spatial and spectral resolution \ion{H}{1} data from Data Release 2 (DR2) of the Galactic Arecibo L-Band Feed Array \ion{H}{1} survey (GALFA-\ion{H}{1}) with the Arecibo 305m radio antenna \citep{2018ApJS..234....2P}. GALFA-\ion{H}{1} has a gridded angular resolution $1'\times1'$ per pixel, a spectral resolution of 0.18 km s$^{-1}$, and a brightness temperature noise of $\simeq$ 40 mK rms per 1 km s$^{-1}$ integrated channel over 13 000 deg$^2$ of sky. The full GALFA-\ion{H}{1} data is separated into three sets: East (denote as A, close to the east of Galactic Plane), North (denote as B, close to the Galactic north pole), and West (denote as C, close to the west of Galactic Plane). Each data set is divided into five individual sub-regions (see Appendix.~\ref{appendx:A} for details about each region). 
	
	For illustration, we take the region A1, which stretches from Right Ascension (R.A.) $\simeq$ 0.0$^\circ$ to 24$^\circ$ and Declination (DEC.) $\simeq$ -1.5$^\circ$ to 37.0$^\circ$ as example. This region spans from galactic latitude b $\simeq -30^\circ$ below the Galactic plane to b $\simeq -60^\circ$ i.e. close to the Galactic south pole. We analysis the \ion{H}{1} data within velocity range -23$km/s$ to 32$km/s$, which contains the main structure of the \ion{H}{1} data.
	
	Following the recipe of VGT-PCA, we show the morphology of plane-of-sky magnetic fields inferred from VGT-PCA with resolution $\simeq 1 ^\circ$ in Fig.~\ref{fig:A1}. We make comparison with the magnetic field morphology inferred from Planck polarization and get AM $=0.77\pm0.02$, which indicates the overall good alignment between VGT-PCA and Planck. The Q, U maps from Planck are smoothed with FWHM=1$^\circ$. The uncertainty is given by the standard error of the mean, i.e. the standard deviation divided by the square root of the sample size.
	
	With an analogy to the Velocity Gradients Technique, the minimum tracing resolution, i.e., the sub-block size, depends on the histogram of gradients orientation. The histogram should be a well defined Gaussian distribution so that the most probable angle can be found out by Gaussian fitting. Once the minimum scale is determined, one can trace the magnetic field in any scale larger than the minimum threshold. In view of this, we test three resolutions (0.25$^\circ$, 0.50$^\circ$, and 1.0$^\circ$) for VGT-PCA and plot the histogram of the relative orientation between the magnetic field predicted by VGT-PCA and the one inferred from Planck polarization in Fig.~\ref{fig:A1}. We see that for all of that three resolutions, the histogram shows a convex Gaussian profile with peak value around zero. 
	
	In view of that there exits deviation between the magnetic field predicted by VGT-PCA and the one inferred from Planck polarization. We consider two possible reasons. One contribution is from the fitting uncertainty of sub-block averaging method. It has been reported that the velocity gradient orientations in a sub-block would form a Gaussian distribution in which the peak of the Gaussian fit reflects the ‘statistical most probable’ magnetic field orientation in this sub–block \citep{2017ApJ...837L..24Y}. As the area of the sampled region increases, the precision of the magnetic field traced through the use of a Gaussian block fit becomes more and more accurate. However, there exits uncertainty in fitting Gaussian distribution and the most probable value of Gaussian distribution has its own standard deviation $\sigma$. Those factors would contribute to the overall uncertainty of gradients calculation \citep{IGs}, which is expected as the systematic error in our calculation.
	The second factor is possibly the uncertainty of polarization measurements. The uncertainty of dust polarimetry becomes significant in the case in which the grains are not aligned. The grain alignment theory suggests that grain alignment is driven mainly by radiative torques, but the grains become misaligned in a number of circumstances \cite{2003JQSRT..79..881L}. For instance, in the absence of sufficiently intense radiation, the orientation of dust grains is random \citep{2007ApJ...669L..77L}. In view of this, we plot the correlation between the AM and polarization percentage $p$ in Fig.~\ref{fig:A1}. We take the average value of AM in each bin of the polarization percentage. We see that the AM is positively proportional to the polarization percentage and gets saturated when $p\ge8\%$. Therefore, the insufficient polarization flux in polarimetry measurements contributes to the deviation between the magnetic field predicted by VGT-PCA and the one inferred from Planck polarization.
	
	Except for the sub-region A1 which is close to the Galactic south pole, we repeat our analysis for the sub-region C1 which spans from galactic latitude b $\simeq 30^\circ$ above the Galactic plane to b $\simeq 45^\circ$. The POS magnetic filed morphology is shown in Fig.~\ref{fig:C1}. The $AM =0.75\pm0.02$ indicates the overall good alignment between VGT-PCA and Planck in C1. We see similar results in terms of the histogram of the relative orientation and the variation of AM concerning the polarization percentage. Moreover, we apply those analyses to all low-latitude regions, i.e., data set A and C, we see all of them shows the good alignment between VGT-PCA and Planck and positive correlation between the alignment and polarization percentage (see Appendix.\ref{appendx:A} for details about each region). Therefore, we conclude that the VGT-PCA shows statistically good performance in terms of characterizing the Galactic dust polarization comparing with the Planck polarization. We see the possibility to correct the issue of insufficient polarization percentage through VGT-PCA.  The resultant morphology of plane-of-sky magnetic fields inferred from Planck polarization and VGT-PCA is shown in Fig.~\ref{fig:PGCA_A}.
	\begin{figure*}
		\centering
		\includegraphics[width=.99\linewidth,height=0.48\linewidth]{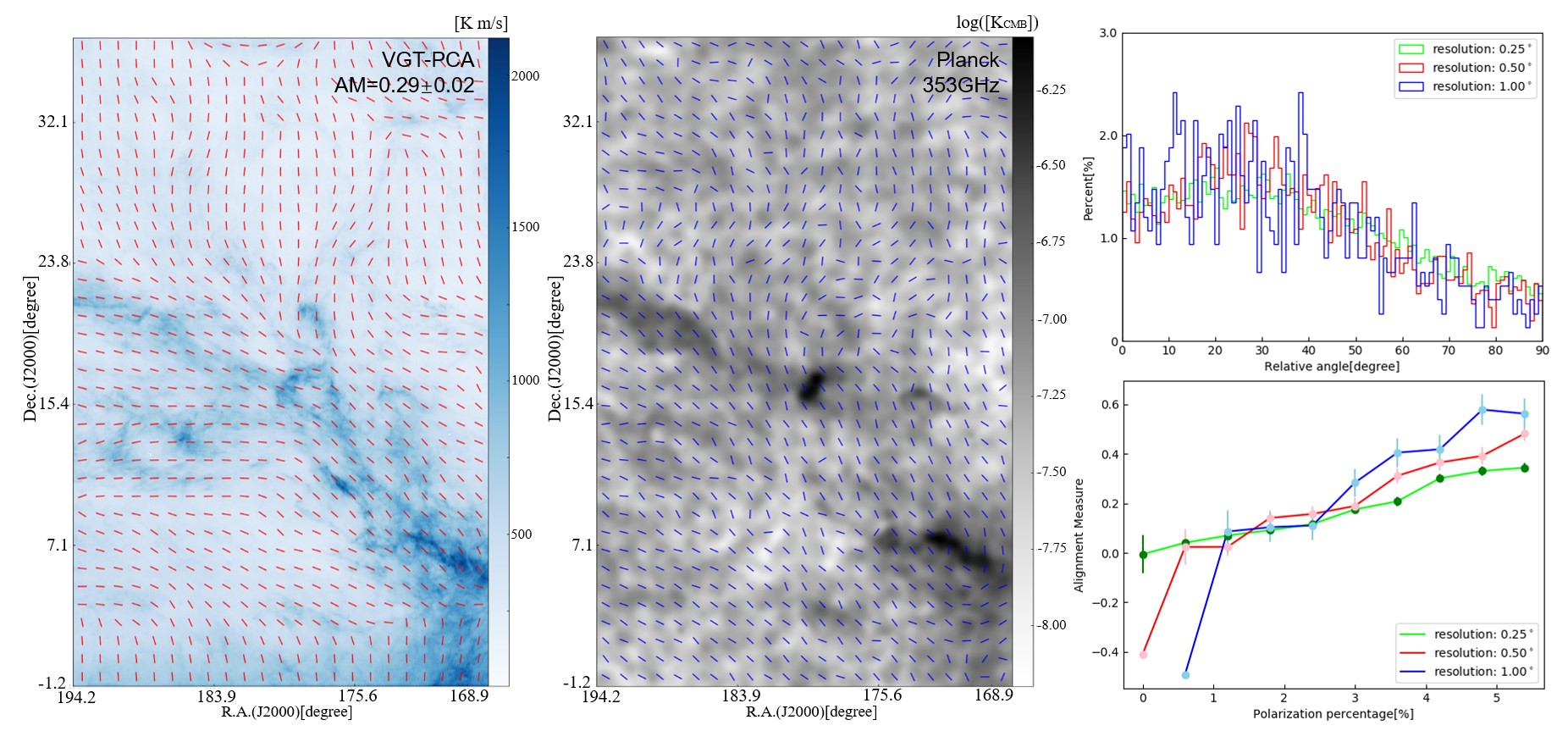}
		\caption{\label{fig:B3} The Morphology of the POS magnetic fields on the sky patch with R.A. from 168.0$^\circ$ to 192.0$^\circ$ and DEC. from -1.2$^\circ$ to 37.1$^\circ$. \textbf{Left:} the magnetic field predicted from VGT-PCA (red segments) with resolution $\simeq 1 ^\circ$. \textbf{Middle:} the magnetic fields inferred from Planck polarization (blue segments). The background color map of VGT-PCA is the GALFA-\ion{H}{1} data integrated  from -41 $km\cdot s^{-1}$ to 14 $km\cdot s^{-1}$. \textbf{Right top:} the histogram of the relative orientation between the magnetic field predicted by VGT-PCA and the one inferred from Planck polarization. \textbf{Right bottom:} the variation of the AM with respect to the polarization percentage.}
	\end{figure*}

	\subsection{Magnetic fields morphology on high Galactic latitude region}
	Based on the advanced understanding of MHD turbulence in ISM, the PCGA shows good ability in tracing the magnetic field morphology at low Galactic regions (see \S.\ref{sec:low latitude}) comparing with Planck polarization. As for high Galactic regions, \ion{H}{1} gas still follows the property of MHD turbulence. Therefore, we expect the VGT-PCA also works in high Galactic regions. Fig.~\ref{fig:B3} takes the sub-region B3, which includes the Galactic north pole, as an example. We do see there is an agreement between VGT-PCA and Planck in relatively high-intensity regions. From the histogram, we find that the peak value of relative orientation between VGT-PCA and Planck locates at $\simeq20^\circ\pm0.79^\circ$, with standard deviation $\simeq 22.5^\circ$ for all three resolutions. While the histogram is more dispersed, the alignment is still increasing with the increment of polarization percentage.
	
\begin{figure*}
		\centering
		\includegraphics[width=.99\linewidth,height=0.65\linewidth]{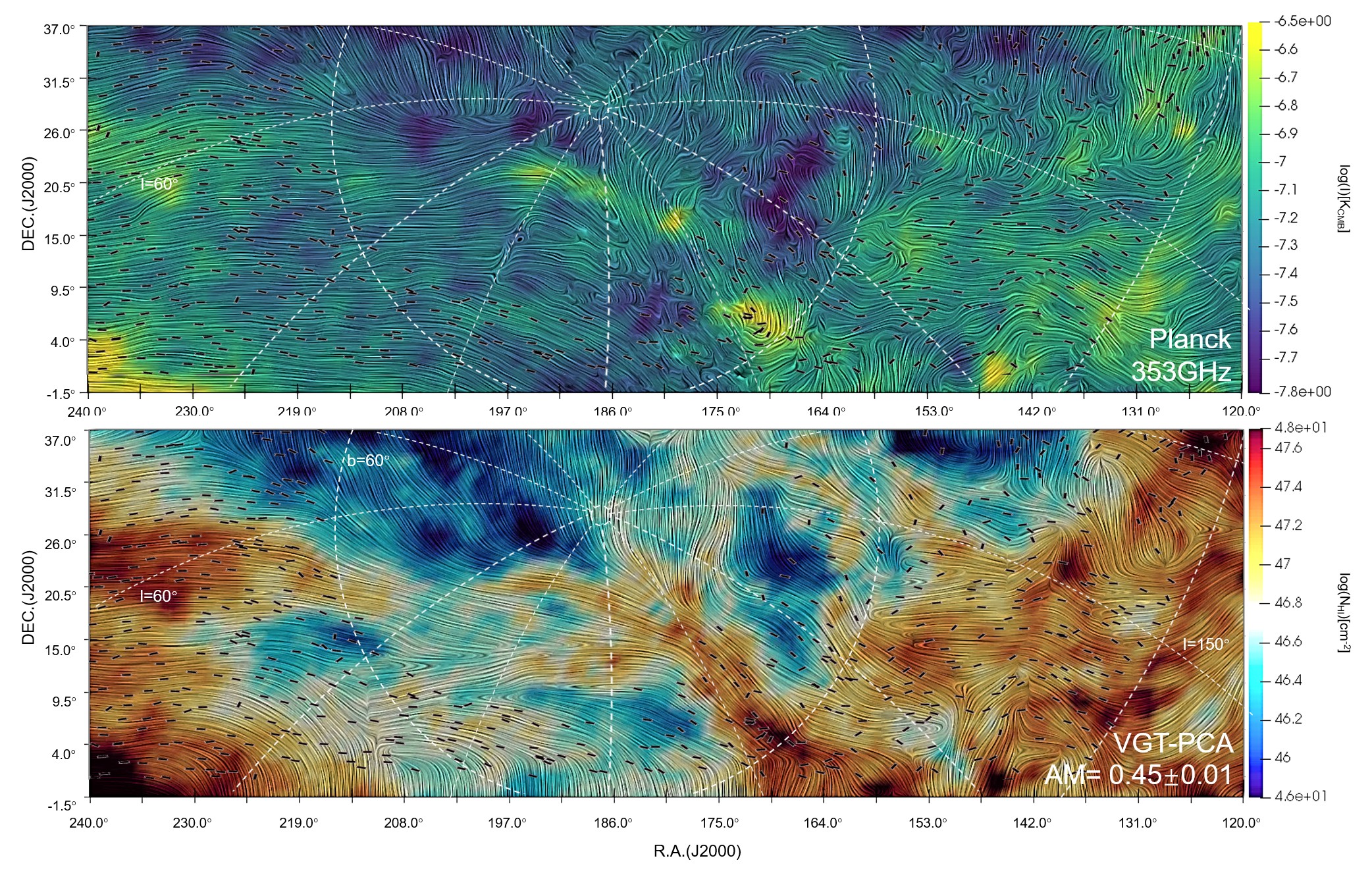}
		\centering
		\caption{\label{fig:PGCA_B} The morphology of the POS magnetic fields at high Galactic latitude regions inferred from Planck polarization (top), and VGT-PCA (bottom). The background color maps are of VGT-PCA is the  \ion{H}{1} column density map integrated from  -90 $km\cdot s^{-1}$ to 90$km\cdot s^{-1}$. The magnetic field is visualized using Line Integral Convolution (LIC). Black pseudo-vectors represent starlight polarization angles obtained from \citet{2014A&A...561A..24B} catalogs.}
	\end{figure*}

	The resultant morphology of plane-of-sky magnetic fields inferred from Planck polarization and VGT-PCA is shown in Fig.~\ref{fig:PGCA_B}. The VGT-PCA in high Galactic latitude regions gives a moderate overall alignment, i.e., AM=$0.45\pm0.01$. The rapidly varying magnetic fields and low signal-to-noise ratio at high Galactic latitude regions are two possible reasons for the misalignment between VGT-PCA and Planck. The polarization percentage $p$ is theoretically determined by the dust column density along the line of sight, and the angle of the mean magnetic field concerning the plane of the sky \citep{2014A&A...571A..11P}. However, at high Galactic latitude regions, the dust emissivity is low \citep{2014A&A...566A..55P}, and the magnetic fields are varying rapidly \citep{2016A&A...594A..10P,2015A&A...576A.104P}. Therefore the polarization fraction is minimal near the Galactic north pole. Besides, the noise level in polarization data is estimated to be higher at the high Galactic latitude region \citep{2015PhRvL.115x1302C}. Those factors obscure the accurate measurement of the Galactic dust polarization and the magnetic field at high Galactic latitude regions. The third possibility of the disagreement might come from the selection of the velocity range in \ion{H}{1} data. The \ion{H}{1} data contains the information of structures in the selected velocity ranges, while polarization is accumulating the information along LOS. Also, we expect the boundary effect will also cause to the disagreement appears near the image's boundary. 
	
\begin{figure}
		\centering
		\includegraphics[width=.95\linewidth,height=0.85\linewidth]{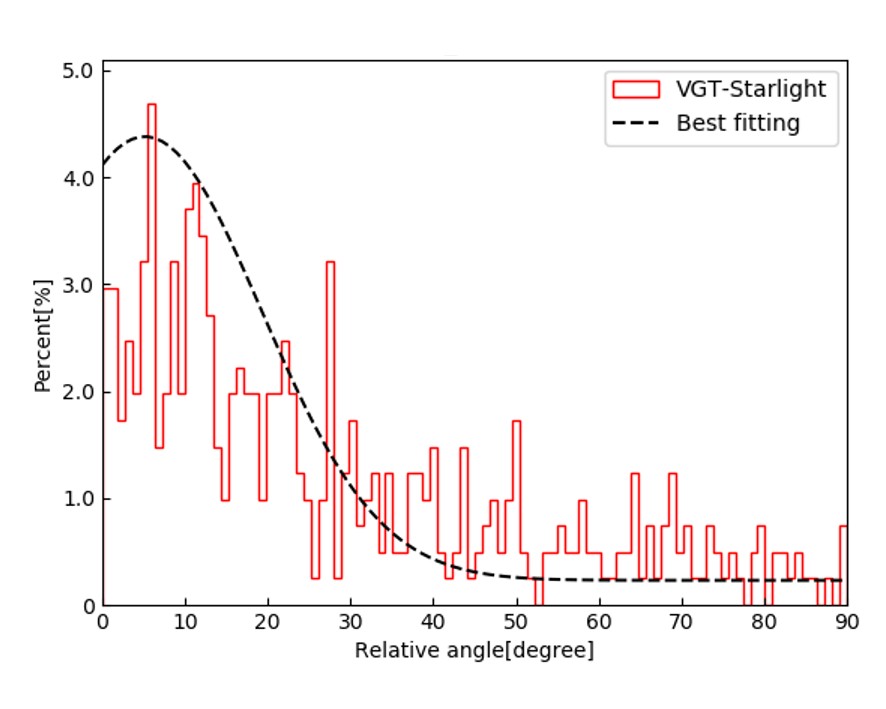}
		\centering
		\caption{\label{fig:star}The histogram of relative angle between the magnetic fields in Fig.~\ref{fig:PGCA_B} derived from VGT-PCA and starlight polarization obtained from \citet{2014A&A...561A..24B} catalogs. The standard
			deviation $\sigma$ is $14.14^\circ$.}
	\end{figure}

	Also, we use the magnetic field inferred from starlight polarization catalogs \citep{2014A&A...561A..24B} as a relative comparison in high Galactic latitude regions ($b>30^\circ$). In Fig.~\ref{fig:star}, we plot the histogram of the relative angle between the polarization from 566 stars and VGT-PCA in corresponding positions. We see that the histogram statistically satisfies a Gaussian distribution, with standard deviation $\sigma=14.4^\circ$ and expectation value $\mu\simeq 5^\circ$. We can, therefore, conclude that the VGT-PCA statistically shows agreement with the starlight polarization. 
	
	\begin{figure}
		\centering
		\includegraphics[width=.95\linewidth,height=0.85\linewidth]{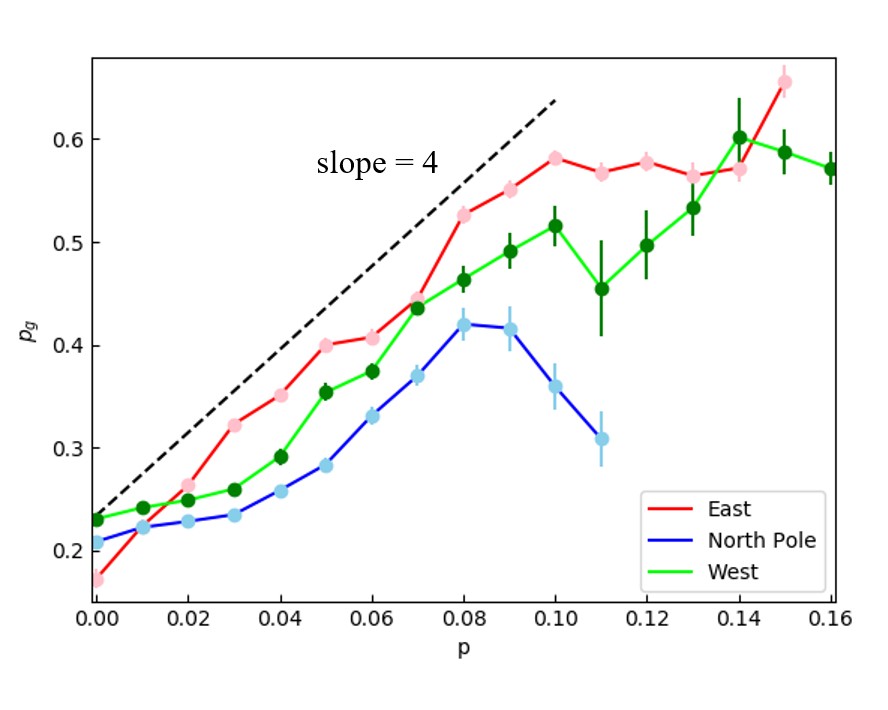}
		\centering
		\caption{\label{fig:Pol_plot}The correlation between the polarization percentage derived from Planck polarization (x-axis, denoted as $p$) and VGT-PCA (y-axis, denoted as $p_g$). }
	\end{figure}
	\begin{figure*}
		\centering
		\includegraphics[width=.99\linewidth,height=0.3\linewidth]{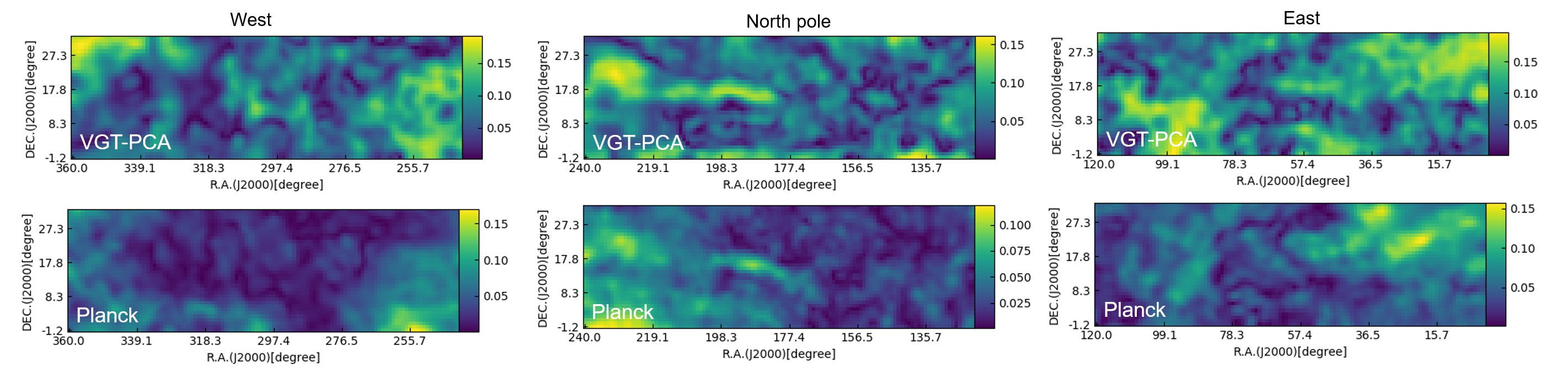}
		\centering
		\caption{\label{fig:Pol}Sky maps of the polarization fraction obtained from VGT-PCA (top) using GALFA-\ion{H}{1} data and Planck 353 GHz dust polarization (bottom) with effective resolution $\simeq 1 ^\circ$.  }
	\end{figure*}
	\subsection{Polarization percentage}
	The polarization percentage $p$ can be derived from Stokes parameters U and Q using the definition in Eq.~\ref{eq.1} and the intensity map from Planck polarization. As for the polarization percentage $p_g$ from pseudo-Stokes parameters U$_g$ and Q$_g$, we use similar definition as $p$ and use the integrated \ion{H}{1} intensiy map. In Fig.~\ref{fig:Pol_plot}, we show the correlation between the polarization percentage derived from Planck polarization (denoted as $p$) and VGT-PCA (denoted as $p_g$) corresponding to the same coordinate. We bin the measured polarization percentage in uniform spaced bins with an interval of $0.01$ . We see that $p_g$ shows a linear correlation with $p$ at low Galactic latitude regions. However, the value of $p_g$ is about four times larger than $p$. We expect the ratio four is the systematic difference between \ion{H}{1} and polarized dust emission data caused by the grain alignment efficiency. 
	
	Generally, it is difficult to model grain alignment efficiency across the sky with different physical conditions. \citet{2018arXiv180706212P} reported a maximum polarization fraction $p_{max}\simeq 22\%$ using 353GHz polarization data. As a result, the dust grain alignment efficiency should be intrinsically similar to $p_{max}$. In Fig.~\ref{fig:Pol}, we see a linear correlation between the polarization percentage predicted by VGT-PCA and Planck polarization, which indicates that the variation of grain alignment efficiency is small across the sky. Fig.~\ref{fig:Pol} gives the sky maps of polarization percentage obtained from VGT-PCA using GALFA-\ion{H}{1} data and Planck 353 GHz dust polarization. We see that VGT-PCA shows similar structures with Planck polarization in terms of polarization percentage. Note this prediction of polarization percentage is incomplete without the knowledge of inclination angle $\gamma$.
	
	\subsection{Decomposition of E/B modes}
	The full sky cosmic microwave background (CMB) polarization field can be decomposed into E and B components that are signatures of distinct physical processes. To decompose the E-mode and B-mode, we modified the template used in \citet{2015PhRvL.115x1302C}:
	\begin{equation}
		\begin{aligned}
			&Q^*(x,y)= p^*\cdot I_{353}(x,y)\cdot\cos(2\psi^*(x,y))\\
			&U^*(x,y)= p^*\cdot I_{353}(x,y)\cdot\sin(2\psi^*(x,y))\\
		\end{aligned}
	\end{equation}
	where $p^*$ is the polarization percentage derived from either VGT-PCA or Planck polarization, $I_{353}$ is the intensity of Planck 353 GHz polarized dust emission, and $\psi^*$ is the polarization angle defined from either VGT-PCA or Planck polarization. In \citet{2015PhRvL.115x1302C}, the polarization percentage is assumed to be ideally unity. However, this assumption is only reasonable over a small patch of sky. For a large patch of sky, it is indispensable to take the polarization percentage into account. 
	
	The derived Stokes maps $Q^*$ and $U^*$ at 1$^\circ$ resolution are converted to HEALPix2 format \citep{2005ApJ...622..759G} with HEALPix resolution of $N_{side}$ = 512, following \citet{2016A&A...586A.141P}. The Stokes $Q^*$ and $U^*$ maps are then decomposed into $C_l^{EE}$ and $C_l^{BB}$ using the $'anafast'$ routine of HEALPix.

\begin{figure}
		\centering
		\includegraphics[width=.95\linewidth,height=0.8\linewidth]{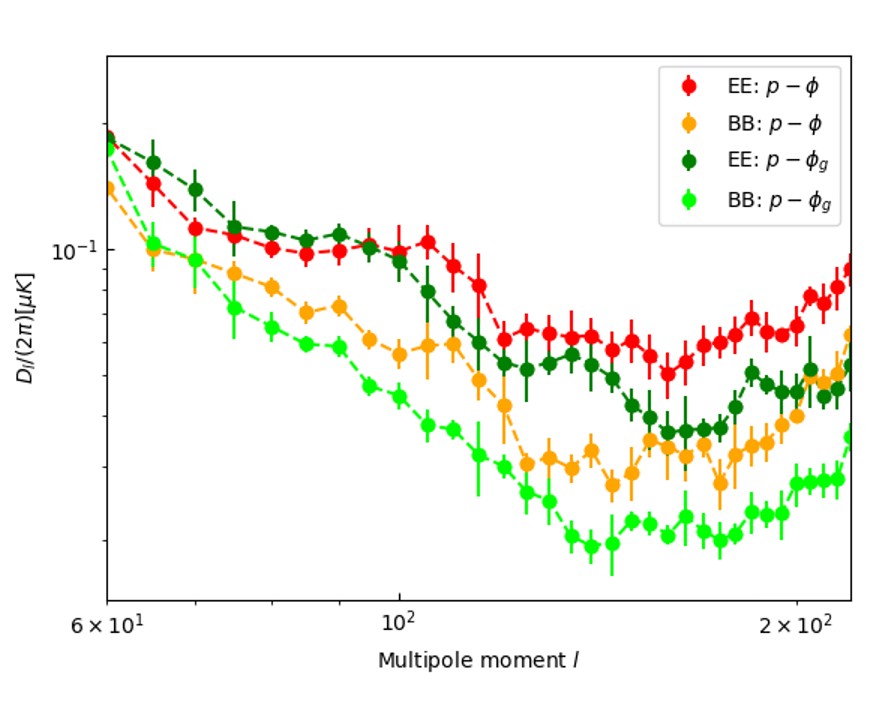}
		\centering
		\caption{\label{fig:EB}The cross-power spectra for the template maps constructed from VGT-PCA and Planck polarization. The y-axis is $D_l=l(l+1)C_l$. The symbol $p$ and $\phi$ indicate that we are using Planck data for the template, while $p_g$ indicates the polarization percentage and polarization angle obtained from VGT-PCA. The spectra are plotted using the full GALFA data set.}
	\end{figure}

	Fig.~\ref{fig:EB} shows the cross-power spectra for the template maps constructed from VGT-PCA and Planck polarization, using the whole set of GALFA-\ion{H}{1} data. The multipole moment ranges from $l=60$ to $l=250$, which is limited by the resolution of input maps. We find that the spectra constructed from VGT-PCA (i.e., the combination of $p$ and $\phi_g$) show a smaller magnitude than the one from Planck polarization (i.e., the combination of $p$ and $\phi$). We expect one possible reason is the contribution from the CMB polarization in Planck HFI data. The discrepancy of magnetic fields traced by two different methods might also cause the deviation. Also, we see that the EE cross-power spectra derived from Planck polarization show a larger amplitude than the BB spectra for both combinations. We find the mean ratio between EE and BB cross-power spectrum derived from VGT-PCA is $0.53\pm0.10$, while it is $0.61\pm0.11$ for Plank polarization. The results coincide with \citet{2016A&A...586A.133P}, which show a systematic difference between the amplitudes of the Galactic dust B- and E-modes $BB/EE\simeq 0.5$.

	\section{Discussion} \label{sec.dis}
	\subsection{Three-dimensional Galactic magnetic fields modeling}
	The detection of primordial B-mode polarization in the CMB is a top topic in cosmology. Although several missions are scheduled to search for the B-mode \citep{2019BAAS...51g.286L,2019arXiv190807495H,2018JCAP...04..023R,2016ARA&A..54..227K,2014JCAP...06..053F,2016A&A...586A.141P,2014PhRvL.112x1101B}, the contamination from the polarized dust emission of the foreground is still an obstacle. To remove the foreground, the model of three-dimensional Galactic magnetic fields, which constrains the LOS structures of the magnetized and dusty ISM, is therefore crucial. However, generally, it is challenging to construct a comprehensive three-dimensional structure of the ISM and the magnetic field orientation along the line-of-sight.
	
	The advanced theories of MHD turbulence \citep{1995ApJ...438..763G,1999ApJ...517..700L} and the statistics of the PPV data \citep{LP00} provide one solution. \citet{1995ApJ...438..763G} and \citet{1999ApJ...517..700L} reveal that the anisotropic turbulence induces velocity fluctuations aligned with its local magnetic fields. \cite{LP00} gave the criterion on selecting thin velocity channels, in which the velocity fluctuations are dominating due to the effect of velocity caustics.  \citet{LY18a} firstly combined these theories to trace the POS magnetic fields using the Velocity Channel Gradients (VChGs).  \citet{2019ApJ...874...25G} extend this concept of thin velocity channels to study the 3D Galactic magnetic fields using \ion{H}{1} data. \citet{2019ApJ...874...25G} initially constructed the 3D magnetic fields in PPV space by adding the velocity gradients in a way similar to Stokes parameter (see Sec.~\ref{sec:method}). Later, \citet{CH19} also implemented this idea in establishing 3D magnetic fields in PPV space. Replying on the galactic rotation curve, \citet{2019ApJ...874...25G} successfully completed the 3D magnetic fields modeling in the Milky Way, comparing with stellar polarization. Similar idea is also expected to be available for the  VGT-PCA. The second way to establish the three-dimensional Galactic magnetic fields requires the knowledge of the inclination angle $\gamma$.
	For example, Eq. \ref{eq.1} indicates that the predicted polarization percentage is overestimated without the picture of three-dimensional magnetic field orientation. 
	
	\subsection{Comparison with Earlier Works}
	\citet{2019arXiv191002226L} demonstrated that Velocity Channel Gradients (VChGs) could be used to produce synthetic maps of dust polarization from \ion{H}{1} data. The VChGs are initially proposed by \citet{LY18a} to trace the magnetic field using thin velocity channels. It selects the data within the LOS velocity range $v_0-\frac{\Delta v}{2} <v< v_0+\frac{\Delta v}{2}$, where $v_0$ is the velocity corresponding to the central peak of the velocity profile along LOS. The velocity channel width $\Delta v$ and the velocity dispersion $\delta v$ satisfy the criterion \citep{LP00}:
	\begin{equation}
		\label{eq1}
		\Delta v^2<\delta v^2
	\end{equation}
	However, in our work, we do not constrain the data within the range $v_0-\frac{\Delta v}{2} <v< v_0+\frac{\Delta v}{2}$, but integrate the pseudo-Stokes parameters obtained in each slice along LOS. \citet{2019arXiv191002226L} limits their work on a sky region, which stretches over R.A. from 215.0$^\circ$ to 265.0$^\circ$ and DEC. from 6.0$^\circ$ to 37.5$^\circ$, to avoid the regions near the Galactic plane and the north Galactic pole. By making synergy with PCA, we extend our prediction of dust polarization over the full GALFA-\ion{H}{1} data set, which covers the Galactic plane, the north Galactic pole, and the south Galactic pole. Our work thus involves several different physical conditions. We further show that the grain alignment efficiency $\eta\simeq 25\%$, and its variation is small across the sky. It indicates that the Galactic magnetic fields cause the variation in polarized dust emission.
	
	\citet{CH19} proposed to predict the dust polarization using \ion{H}{1} in an alternatively way. Their work is based on the Rolling Hough Transform (RHT, see \citealt{2014ApJ...789...82C}), which requires linear structures in ISM. Comparing with RHT, the VGT-PCA technique is parameter-free while RHT requires three parameters as inputs: a smoothing kernel diameter (DK), window diameter (DW), and intensity threshold (Z) \citep{2014ApJ...789...82C,2015PhRvL.115x1302C}. \citet{CH19} predicted dust polarization across the full sky region using HI4PI data. However,  \citet{CH19} interpreted that the predominantly physical nature of the \ion{H}{1} structures in thin velocity channel maps comes from Cold Neutral Media (CNM), without the consideration of the velocity caustics effect \citep{LP00}. It thus questions (i) the physical nature of "\ion{H}{1} fibers" and (ii) the distribution of CNM in the Galaxy. For the latter, \citet{2018A&A...619A..58K} showed that the CNM concentrates on the region near the Galactic plane, i.e., $|b|<30^\circ$, instead of the full sky. 
	
	\subsection{Application to Giant Molecular Clouds}
	The approach of gradients technique has been successfully tested to trace the local magnetic field from absorbing media for the case of $^{13}$CO emission with different abundances and densities \citep{2019MNRAS.483.1287G}, while \citet{2019ApJ...873...16H} numerically showed the availability of the gradient in tracing magnetic fields using synthetic molecular line maps of CO isotopologue with different optical depths. Thereafter, in observation, \citet{survey} successfully applied the gradients technique to five low mass star forming regions using $^{13}$CO as molecular tracer, while \citet{velac} expanded the technique into seven different molecular tracers in Giant Molecular Clouds Vela C. Importantly, we see that the magnetic field morphology obtained from VGT-PCA in the region corresponding to molecular clouds Taurus and Perseus (see Appendix.~\ref{appendx:A}, Fig.~\ref{fig:A45}) agrees with the results from Planck polarization and the Velocity Gradients Technique in \citet{survey}. Therefore, we expect the VGT-PCA is also applicable to molecular clouds.
	
	Due to the position of the solar system within the Galactic disk, the line of sight inevitably crosses more than one molecular cloud. It is therefore impossible to use far-infrared polarimetry to study the local magnetic fields in most molecular clouds. Fortunately, VGT-PCA show advantages in dealing with multi-clouds issues. Theoretically both VGT-PCA and dust polarization represent the direction of the projected magnetic field by accumulating the information along the line of sight. However, spectroscopic data often sample different regions of the molecular clouds than dust polarization. Spectroscopic data cubes can better identify the molecular cloud in velocity space with respect to the polarization data. The background/foreground signals are therefore less important in the spectroscopic data cubes. VGT-PCA using molecular tracers would provide the information of the local magnetic field in the cloud, while polarization accumulates the information along the line of sight. Combining the magnetic field traced from \ion{H}{1} data, CO or other molecular tracers data, and polarized dust emission data, we see the possibility to figure out the contribution from the Galactic foreground to polarized dust emission on molecular clouds.
	
	\subsection{Application within synchrotron polarization}
	The VGT-PCA is one of the techniques that employ the properties of MHD turbulence to study magnetic fields. We expect the application of the VGT-PCA is not limited to the molecular emission line data. \cite{2018ApJ...855...72L}, \cite{2018ApJ...865...59L}, and \citet{2019MNRAS.tmp.1175Z} numerically show that the Synchrotron Intensity Gradients (SIGs) and Synchrotron Polarization Gradients (SPGs) can be used to trace the magnetic field component parallel to the LOS, and even the three-dimensional magnetic field morphology. Since the foundation of SIGs and SPGs is still the anisotropy of MHD turbulence, we expect the VGT-PCA is applicable to synchrotron data as well.
	
	\section{Conclusion} \label{sec:con}
	The inflationary gravitational wave B-mode polarization is contaminated by the Galactic foreground polarization arising from complicated interstellar magnetic fields. In this work, we explore a new way to trace the magnetic field and make prediction to the polarized dust emission using neutral hydrogen data. To summary:
	\begin{enumerate}
		\item We develop the VGT-PCA as a new tool in characterizing the Galactic dust polarization using only neutral hydrogen data, by making synergy of VGT and PCA. In particular:
		\begin{enumerate}
			\item We apply the VGT-PCA to all high-resolution neutral hydrogen data from the GALFA-\ion{H}{1} survey and make a comparison with the Planck polarized dust emission and stellar polarization data. 
			\item The VGT-PCA shows statistically good agreement with the Planck polarization and stellar polarization in term of magnetic field tracing.
			\item We find that the alignment between the VGT-PCA and Planck polarization is positively proportional to the polarization percentage. We conclude the insufficient polarization flux is one factor leads to the misalignment of the VGT-PCA and Planck polarization.
		\end{enumerate}    
		\item  We find the polarization percentage predicted by VGT-PCA using GALFA-\ion{H}{1} data and Planck polarization is linearly correlated with a factor 0.25 and its variation is small across the sky. 
		\item The full Galactic foreground templates can be constructed by using the Stokes parameter predicted by VGT-PCA. We show that the mean ratio between EE and BB cross-power spectrum derived from VGT-PCA is $0.53\pm0.10$, which is in agreement with the resuts from Planck measurement. 
		\item We claim the VGT-PCA as a modification of the VGT that is advantageous in
		studying the magnetic field morphology and predicting the contribution from the Galactic foreground.
	\end{enumerate}
	
	\noindent
	{\bf Acknowledgment}  {A.L. acknowledges the support of the NSF grant AST 1715754, and 1816234. K.H.Y. acknowledges the support of the NSF grant AST 1816234. Y.H. acknowledges the support of the NASA TCAN 144AAG1967. This publication utilizes data from Galactic ALFA HI (GALFA HI) survey data set obtained with the Arecibo L-band Feed Array (ALFA) on the Arecibo 305m telescope. The Arecibo Observatory is operated by SRI International under a cooperative agreement with the National Science Foundation (AST-1100968), and in alliance with Ana G. Méndez-Universidad Metropolitana, and the Universities Space Research Association. The GALFA HI surveys have been funded by the NSF through grants to Columbia University, the University of Wisconsin, and the University of California. We acknowledge the use of NERSC TCAN computer time allocation.}
	\software{Julia \citep{2012arXiv1209.5145B}, HEALPix \citep{2005ApJ...622..759G}, Paraview \citep{ayachit2015paraview}}
	
	\newpage
	\appendix
	\section{GALFA-\ion{H}{1} data selection}
	\label{appendx:A}
\begin{figure*}
		\centering
		\subfigure[A2]{
			\includegraphics[width=.99\linewidth,height=0.5\linewidth]{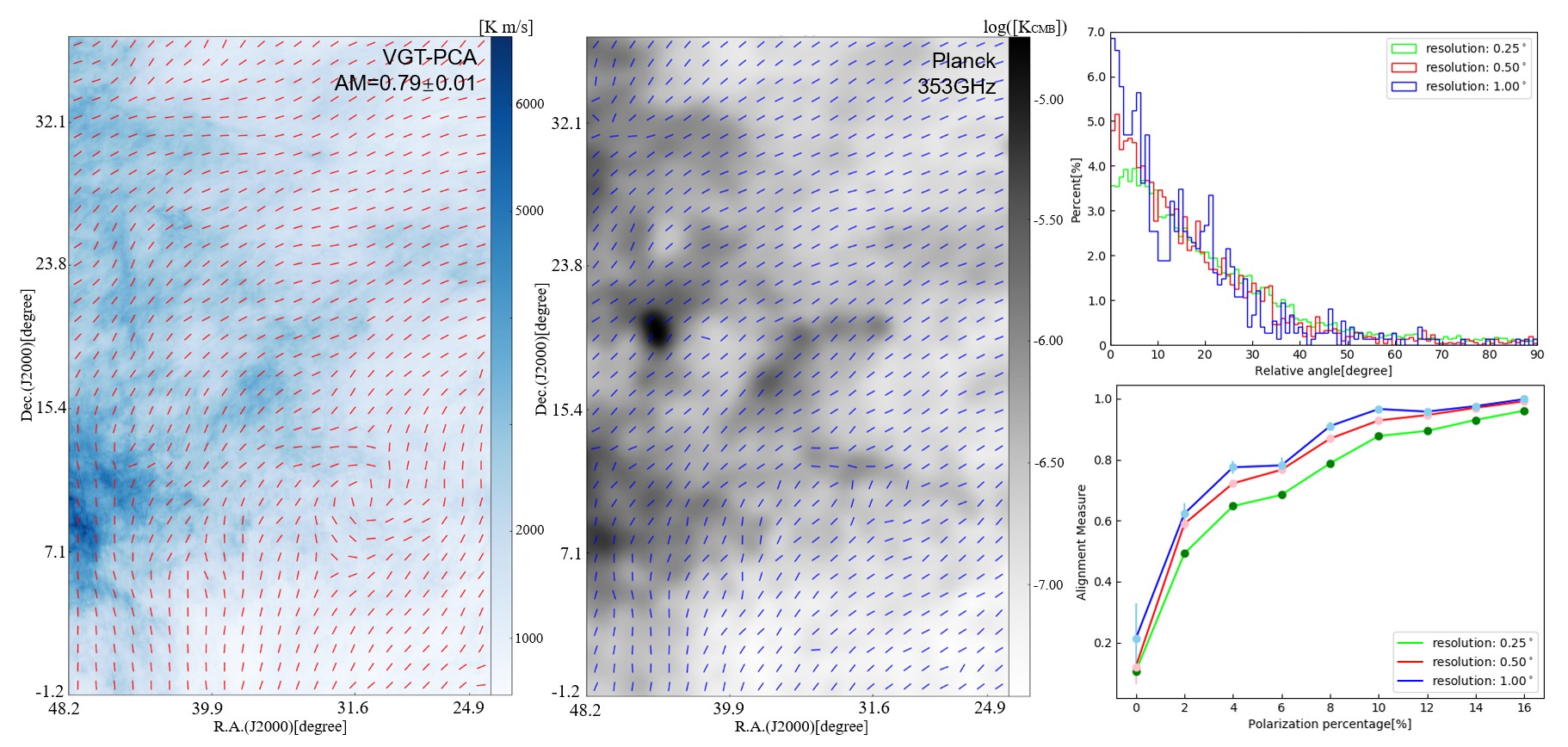}}
		\centering
		\subfigure[A3]{
			\includegraphics[width=.99\linewidth,height=0.5\linewidth]{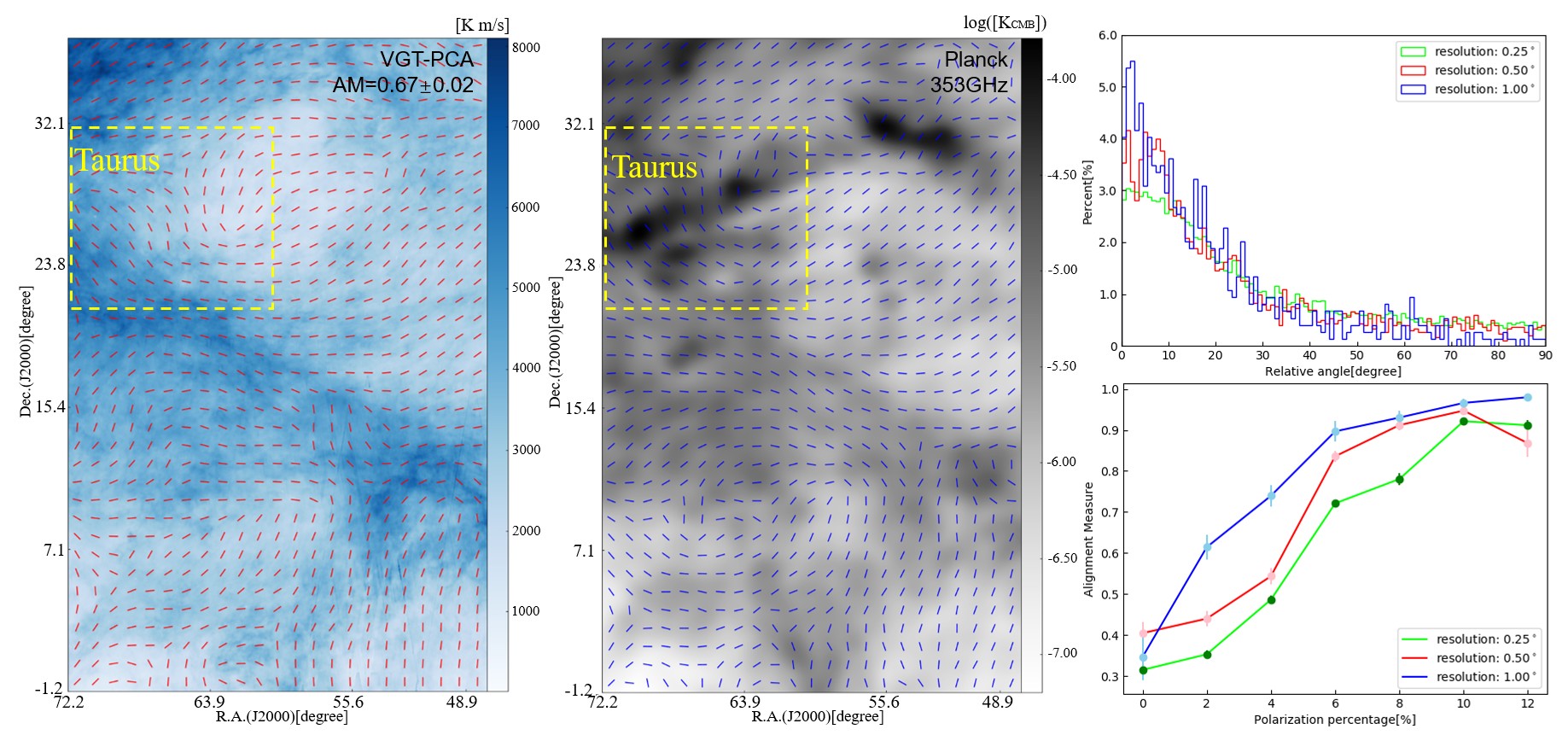}}
		\caption{\label{fig:A23} The Morphology of the POS magnetic fields on the sky patch with \textbf{(a):} R.A. from 24.0$^\circ$ to 48.0$^\circ$ and DEC. from -1.2$^\circ$ to 37.1$^\circ$, and \textbf{(b):} R.A. from 48.0$^\circ$ to 72.0$^\circ$ and DEC. from -1.2$^\circ$ to 37.1$^\circ$. \textbf{Left:} the magnetic field predicted from VGT-PCA (red segments) with resolution $\simeq 1 ^\circ$. \textbf{Middle:} the magnetic fields inferred from Planck polarization (blue segments). \textbf{Right top:} the histogram of the relative orientation between the magnetic field predicted by VGT-PCA and the one inferred from Planck polarization. \textbf{Right bottom:} the variation of the AM with respect to the polarization percentage.}
	\end{figure*}

\begin{figure*}
		\centering
		\subfigure[A4]{
			\includegraphics[width=.99\linewidth,height=0.5\linewidth]{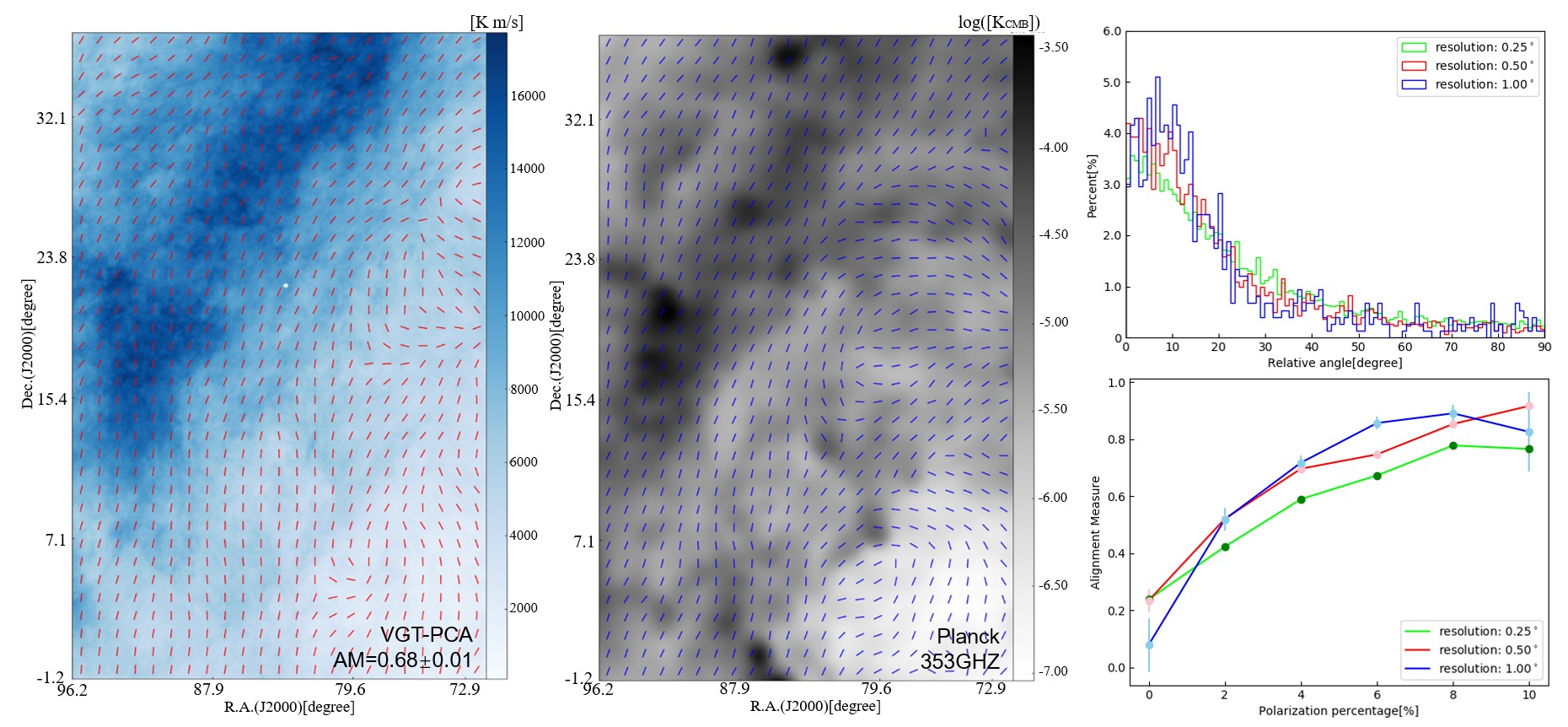}}
		\centering
		\subfigure[A5]{
			\includegraphics[width=.99\linewidth,height=0.5\linewidth]{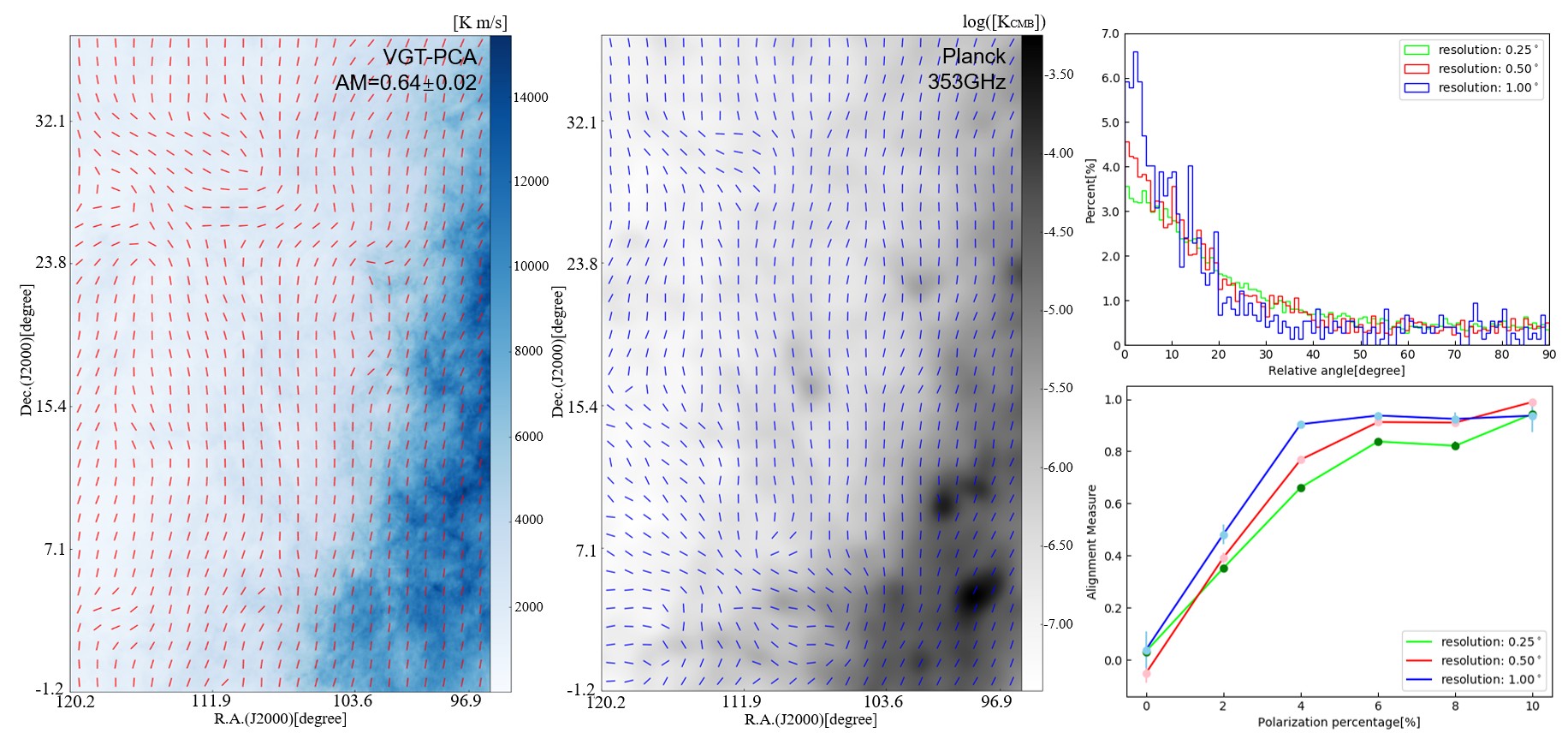}}
		\caption{\label{fig:A45} The Morphology of the POS magnetic fields on the sky patch with \textbf{(a):} R.A. from 72.0$^\circ$ to 96.0$^\circ$ and DEC. from -1.2$^\circ$ to 37.1$^\circ$, and \textbf{(b):} R.A. from 96.0$^\circ$ to 120.0$^\circ$ and DEC. from -1.2$^\circ$ to 37.1$^\circ$. \textbf{Left:} the magnetic field predicted from VGT-PCA (red segments) with resolution $\simeq 1 ^\circ$. \textbf{Middle:} the magnetic fields inferred from Planck polarization (blue segments). \textbf{Right top:} the histogram of the relative orientation between the magnetic field predicted by VGT-PCA and the one inferred from Planck polarization. \textbf{Right bottom:} the variation of the AM with respect to the polarization percentage.}
	\end{figure*}

\begin{figure*}
		\centering
		\subfigure[C2]{
			\includegraphics[width=.99\linewidth,height=0.5\linewidth]{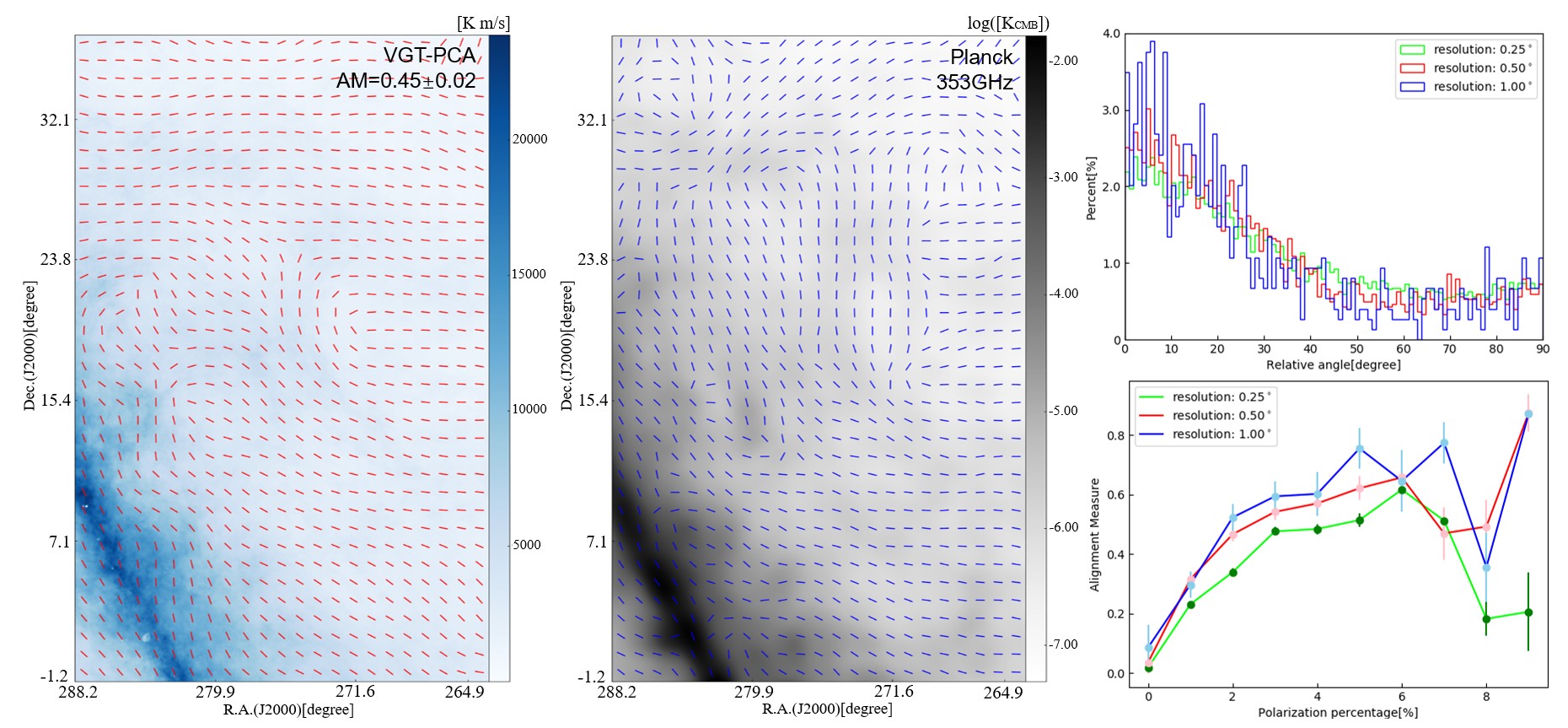}}
		\centering
		\subfigure[C3]{
			\includegraphics[width=.99\linewidth,height=0.5\linewidth]{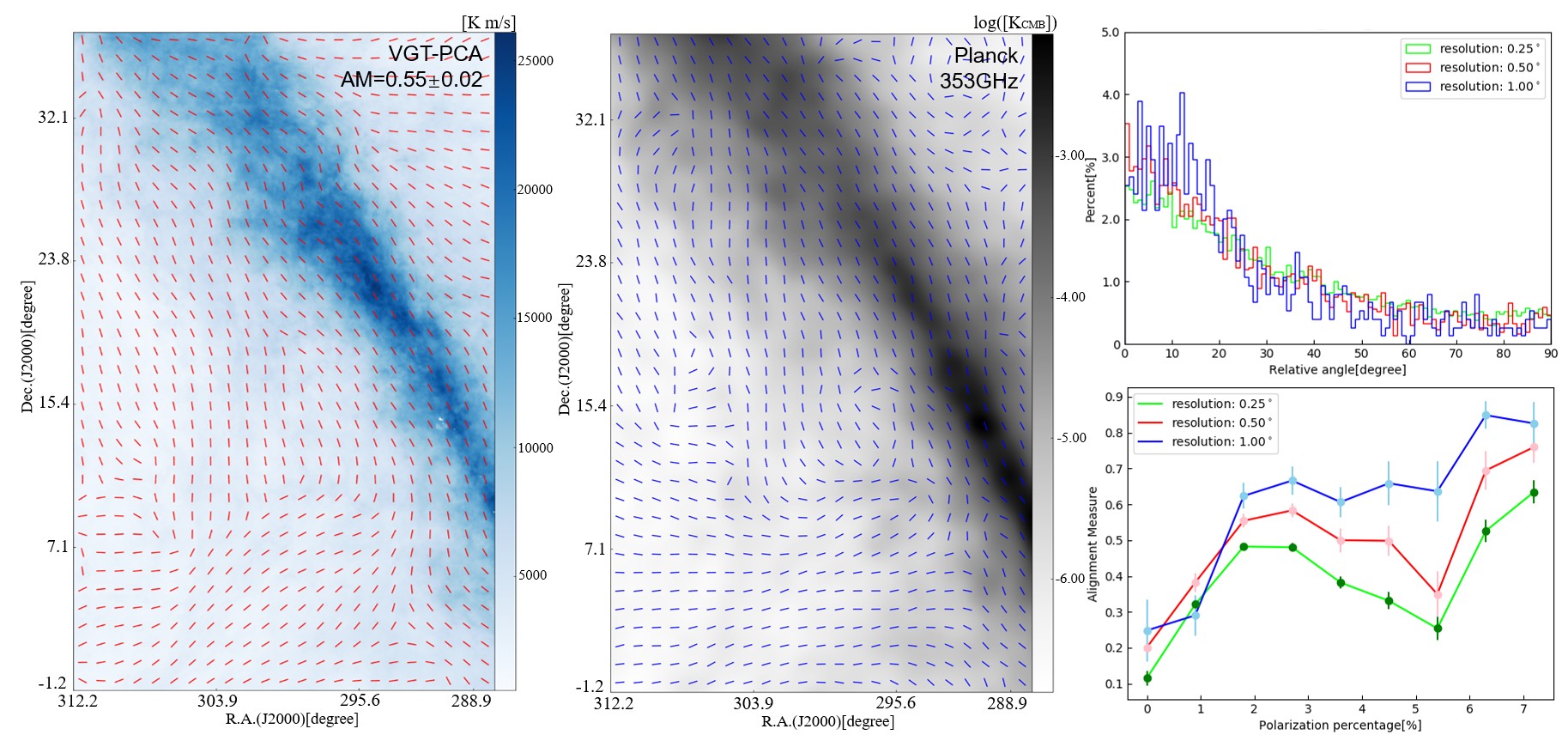}}
		\caption{\label{fig:C23} The Morphology of the POS magnetic fields on the sky patch with \textbf{(a):} R.A. from 264.0$^\circ$ to 288.0$^\circ$ and DEC. from -1.2$^\circ$ to 37.1$^\circ$, and \textbf{(b):} R.A. from 288.0$^\circ$ to 312.0$^\circ$ and DEC. from -1.2$^\circ$ to 37.1$^\circ$. \textbf{Left:} the magnetic field predicted from VGT-PCA (red segments) with resolution $\simeq 1 ^\circ$. \textbf{Middle:} the magnetic fields inferred from Planck polarization (blue segments). \textbf{Right top:} the histogram of the relative orientation between the magnetic field predicted by VGT-PCA and the one inferred from Planck polarization. \textbf{Right bottom:} the variation of the AM with respect to the polarization percentage.}
	\end{figure*}

\begin{figure*}
		\centering
		\subfigure[C4]{
			\includegraphics[width=.99\linewidth,height=0.5\linewidth]{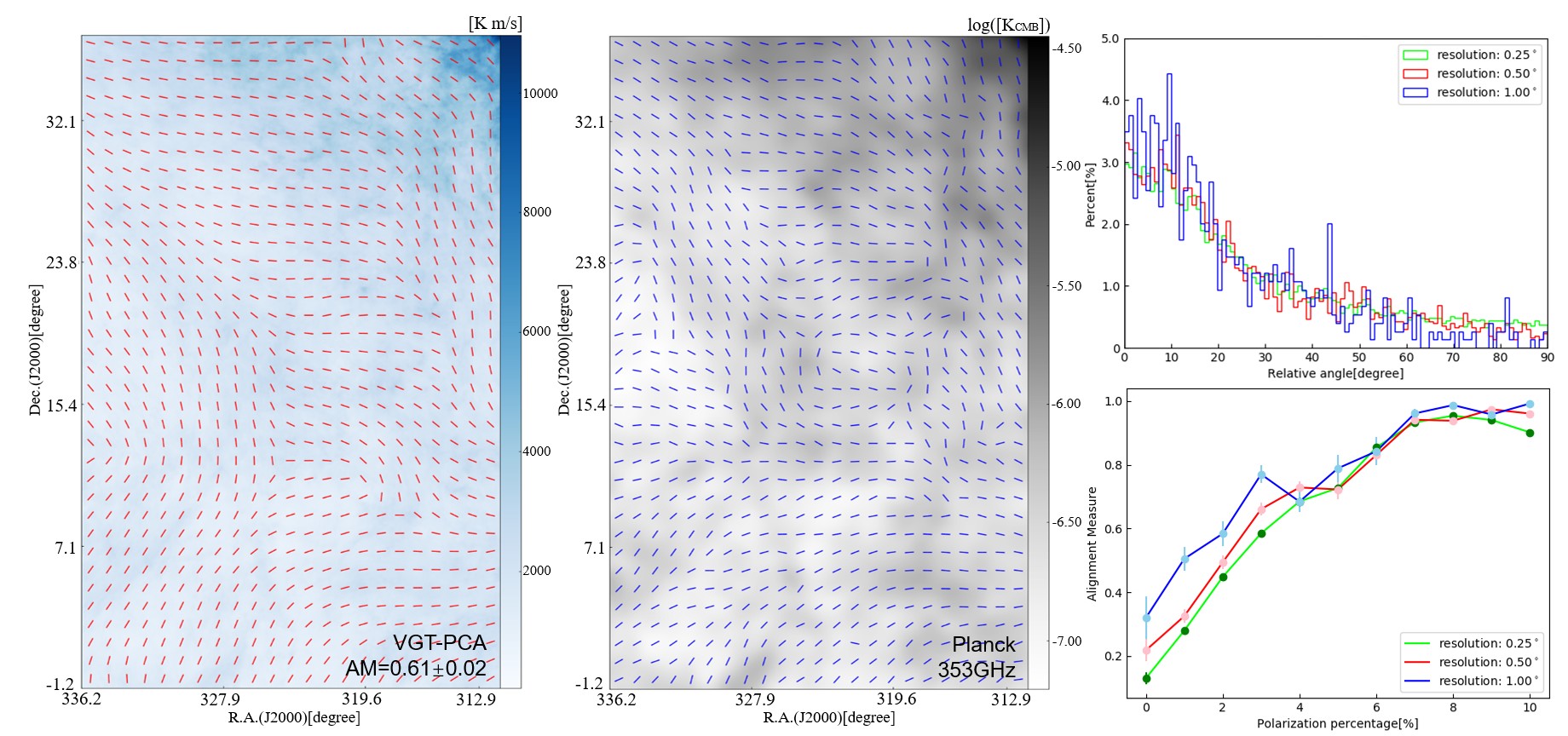}}
		\centering
		\subfigure[C5]{
			\includegraphics[width=.99\linewidth,height=0.5\linewidth]{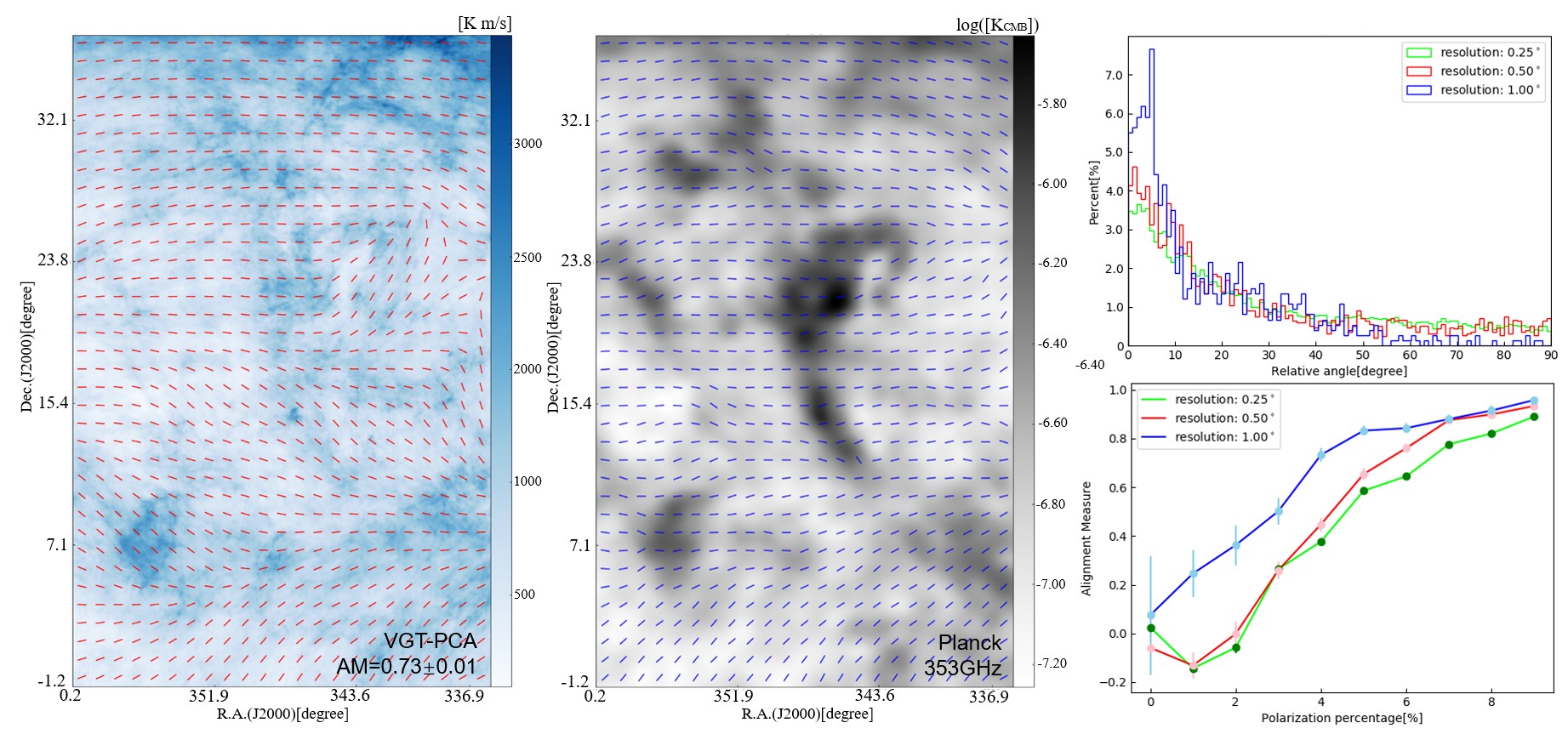}}
		\caption{\label{fig:C45} The Morphology of the POS magnetic fields on the sky patch with \textbf{(a):} R.A. from 312.0$^\circ$ to 336.0$^\circ$ and DEC. from -1.2$^\circ$ to 37.1$^\circ$, and \textbf{(b):} R.A. from 336.0$^\circ$ to 360.0$^\circ$ and DEC. from -1.2$^\circ$ to 37.1$^\circ$. \textbf{Left:} the magnetic field predicted from VGT-PCA (red segments) with resolution $\simeq 1 ^\circ$. \textbf{Middle:} the magnetic fields inferred from Planck polarization (blue segments). \textbf{Right top:} the histogram of the relative orientation between the magnetic field predicted by VGT-PCA and the one inferred from Planck polarization. \textbf{Right bottom:} the variation of the AM with respect to the polarization percentage. }
	\end{figure*}

\begin{figure*}
		\centering
		\subfigure[B1]{
			\includegraphics[width=.99\linewidth,height=0.5\linewidth]{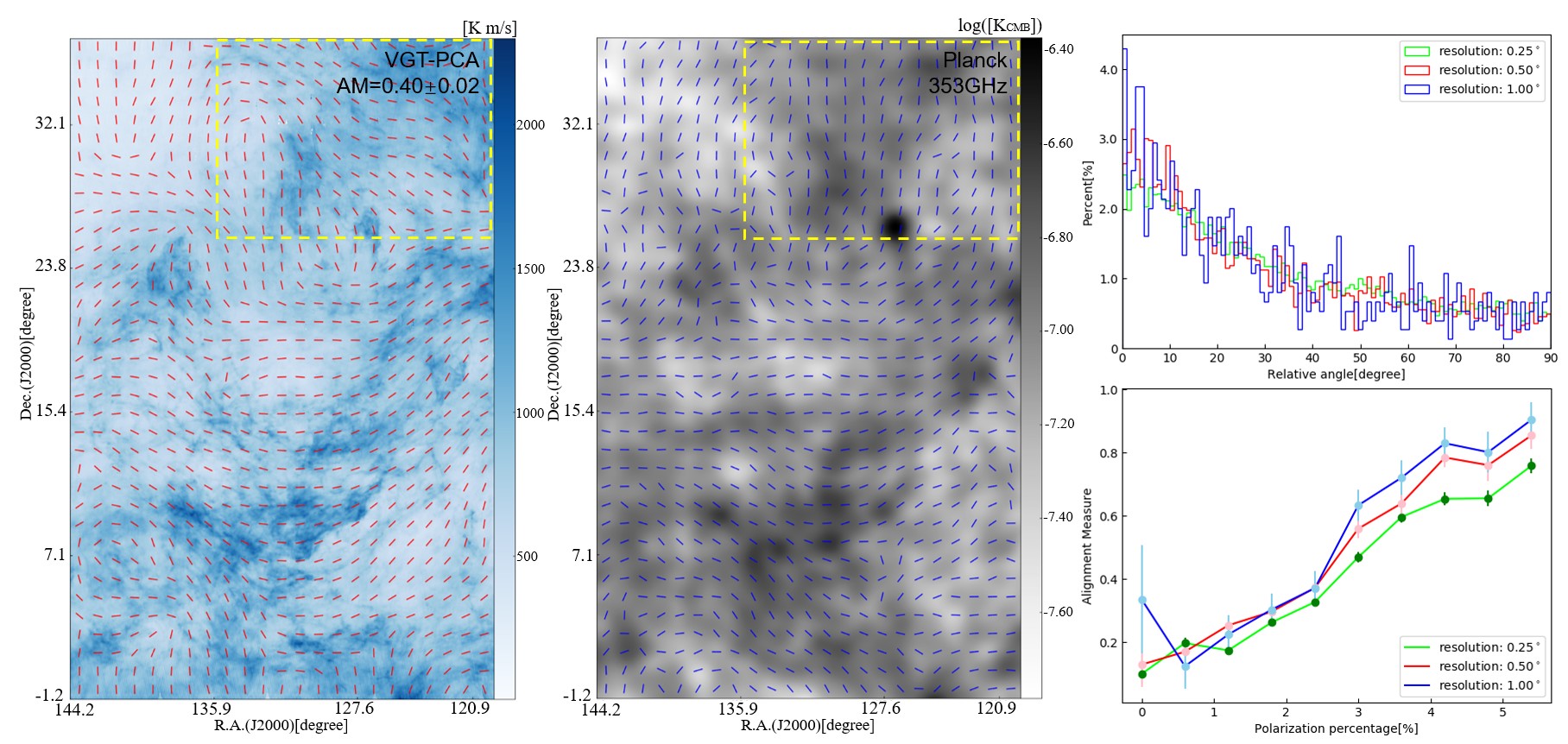}}
		\centering
		\subfigure[B2]{
			\includegraphics[width=.99\linewidth,height=0.5\linewidth]{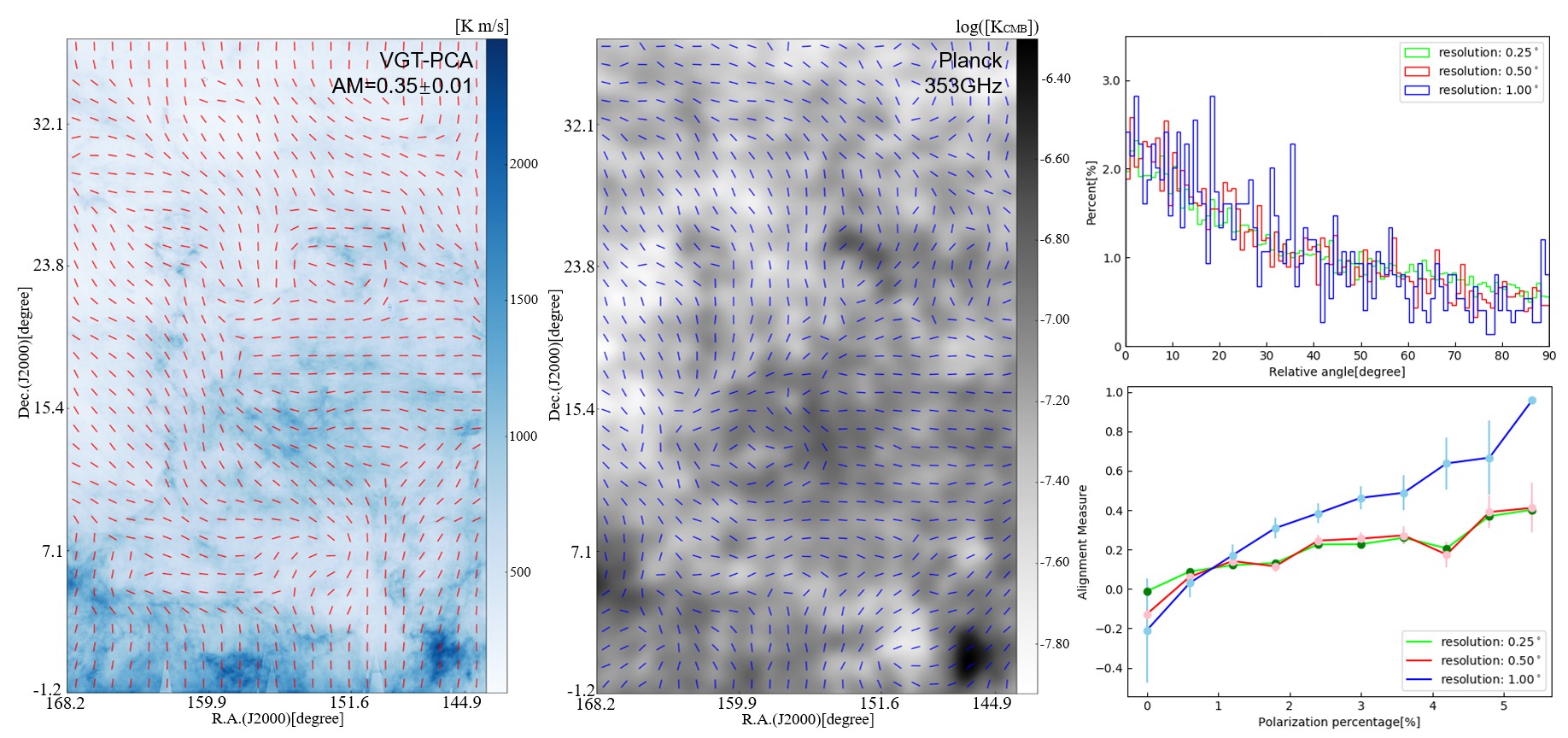}}
		\caption{\label{fig:B12} The Morphology of the POS magnetic fields on the sky patch with \textbf{(a):} R.A. from 120.0$^\circ$ to 144.0$^\circ$ and DEC. from -1.2$^\circ$ to 37.1$^\circ$, and \textbf{(b):} R.A. from 144.0$^\circ$ to 360.0$^\circ$ and DEC. from -1.2$^\circ$ to 37.1$^\circ$. \textbf{Left:} the magnetic field predicted from VGT-PCA (red segments) with resolution $\simeq 1 ^\circ$. \textbf{Middle:} the magnetic fields inferred from Planck polarization (blue segments). \textbf{Right top:} the histogram of the relative orientation between the magnetic field predicted by VGT-PCA and the one inferred from Planck polarization. \textbf{Right bottom:} the variation of the AM with respect to the polarization percentage.}
	\end{figure*}

\begin{figure*}
		\centering
		\subfigure[B4]{
			\includegraphics[width=.99\linewidth,height=0.5\linewidth]{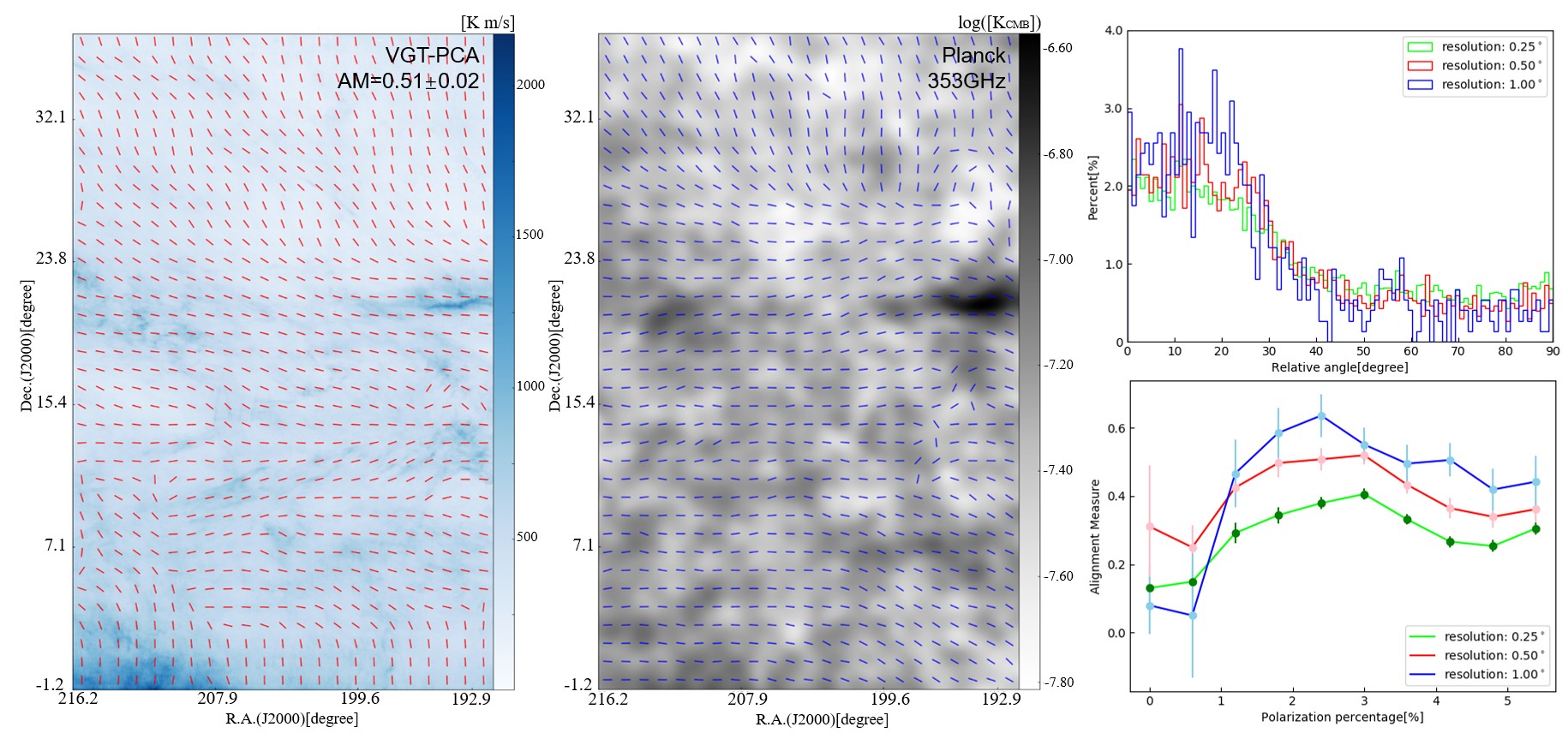}}
		\centering
		\subfigure[B5]{
			\includegraphics[width=.99\linewidth,height=0.5\linewidth]{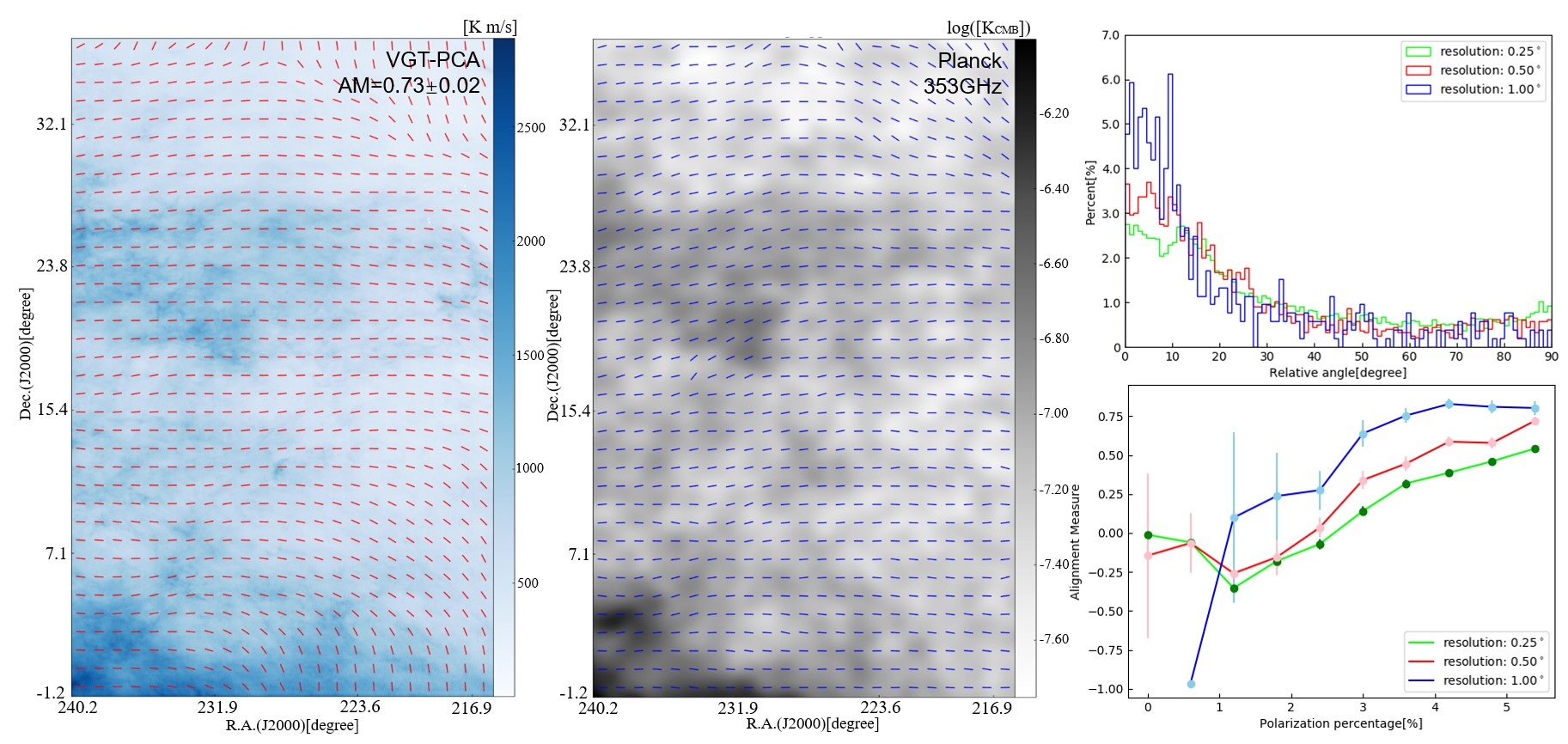}}
		\caption{\label{fig:B45} The Morphology of the POS magnetic fields on the sky patch with \textbf{(a):} R.A. from 192.0$^\circ$ to 216.0$^\circ$ and DEC. from -1.2$^\circ$ to 37.1$^\circ$, and \textbf{(b):} R.A. from 216.0$^\circ$ to 240.0$^\circ$ and DEC. from -1.2$^\circ$ to 37.1$^\circ$. \textbf{Left:} the magnetic field predicted from VGT-PCA (red segments) with resolution $\simeq 1 ^\circ$. \textbf{Middle:} the magnetic fields inferred from Planck polarization (blue segments). \textbf{Right top:} the histogram of the relative orientation between the magnetic field predicted by VGT-PCA and the one inferred from Planck polarization. \textbf{Right bottom:} the variation of the AM with respect to the polarization percentage.}
	\end{figure*}

	\begin{table*}[h]
		\centering
		\label{tab:galfa}
		\begin{tabular}{| c | c | c | c | c | c | c|}
			\hline
			Data Set & Sub-region & R.A.(J2000) & DEC.(J2000)  & Velocity Range & AM & $\mu$\\ \hline \hline
			& A1 & [0.0$^\circ$, 24.0$^\circ$] & [-1.2$^\circ$, 37.1$^\circ$] &  [-32km/s, 23km/s] & 0.77$\pm$0.02 & 1.21$^\circ\pm$0.72$^\circ$\\ 
			& A2 & [24.0$^\circ$, 48.0$^\circ$] & [-1.2$^\circ$, 37.1$^\circ$] & [-32km/s, 23km/s] & 0.79$\pm$0.01 & 2.08$^\circ\pm$0.51$^\circ$\\
			East& A3 & [48.0$^\circ$, 72.0$^\circ$] & [-1.2$^\circ$, 37.1$^\circ$] & [-32km/s, 23km/s]& 0.67$\pm$0.02 & 1.88$^\circ\pm$0.67$^\circ$\\ 
			& A4 & [72.0$^\circ$, 96.0$^\circ$] & [-1.2$^\circ$, 37.1$^\circ$] & [-32km/s, 41km/s]&  0.68$\pm$0.01 & 6.62$^\circ\pm$0.70$^\circ$\\
			& A5 & [96.0$^\circ$, 120.0$^\circ$] & [-1.2$^\circ$, 37.1$^\circ$] & [-23km/s, 32km/s]& 0.64$\pm$0.02 & 2.14$^\circ\pm$0.79$^\circ$\\\hline
			& B1 & [120.0$^\circ$, 144.0$^\circ$] & [-1.2$^\circ$, 37.1$^\circ$] & [-23km/s, 32km/s]& 0.40$\pm$0.02& 0.54$^\circ\pm$0.87$^\circ$\\
			& B2 & [144.0$^\circ$, 168.0$^\circ$] & [-1.2$^\circ$, 37.1$^\circ$] & [-41km/s, 14km/s]& 0.35$\pm$0.01& 13.82$^\circ\pm$0.83$^\circ$\\
			North Pole& B3 & [168.0$^\circ$, 192.0$^\circ$] & [-1.2$^\circ$, 37.1$^\circ$] & [-41km/s, 14km/s]& 0.29$\pm$0.02 & 20.05$^\circ\pm$0.79$^\circ$\\
			& B4 & [192.0$^\circ$, 216.0$^\circ$] & [-1.2$^\circ$, 37.1$^\circ$] & [-32km/s, 23km/s] & 0.51$\pm$0.02& 14.91$^\circ\pm$0.77$^\circ$\\
			& B5 & [216.0$^\circ$, 240.0$^\circ$] & [-1.2$^\circ$, 37.1$^\circ$] & [-32km/s, 23km/s]& 0.73$\pm$0.02& 4.05$^\circ\pm$0.78$^\circ$\\\hline
			& C1 & [240.0$^\circ$, 264.0$^\circ$] & [-1.2$^\circ$, 37.1$^\circ$] & [-23km/s, 32km/s]& 0.75$\pm$0.02& 1.99$^\circ\pm$0.65$^\circ$\\
			& C2 & [264.0$^\circ$, 288.0$^\circ$] & [-1.2$^\circ$, 37.1$^\circ$] & [-23km/s, 41km/s]& 0.45$\pm$0.02& 5.66$^\circ\pm$0.88$^\circ$\\
			West& C3& [288.0$^\circ$, 312.0$^\circ$] & [-1.2$^\circ$, 37.1$^\circ$] & [-23km/s, 32km/s]& 0.55$\pm$0.02& 7.89$^\circ\pm$0.76$^\circ$\\
			& C4 & [312.0$^\circ$, 336.0$^\circ$] & [-1.2$^\circ$, 37.1$^\circ$] & [-32km/s, 23km/s]& 0.61$\pm$0.02& 3.56$^\circ\pm$0.68$^\circ$\\
			& C5 & [336.0$^\circ$, 360.0$^\circ$] & [-1.2$^\circ$, 37.1$^\circ$] & [-32km/s, 23km/s]& 0.73$\pm$0.01& 3.81$^\circ\pm$0.58$^\circ$\\
			\hline
		\end{tabular}
		\caption{Description of GALFA-\ion{H}{1} data set used in this work. The full GALFA-\ion{H}{1} data set is separated into three parts. $\mu$ is the expectation value of the histogram of the relative alignment between VGT-PCA and Planck 353GHz dust polarization, with resolution $\sim 1^\circ$. The uncertainties on AM and $\mu$ are given by the standard error of the mean, i.e., the standard deviation divided by the square root of the sample size.}
	\end{table*}
	In this work, the ensemble GALFA-\ion{H}{1} data is separated into three data sets: East (A), North pole(B), and West (C). Each data set is divided into five individual sub-regions. The velocity range of each data set is selected, such that the majority of the Galactic neutral hydrogen gas is included. Tab.\ref{tab:galfa} gives the description of GALFA-\ion{H}{1} data set used in this work. The uncertainty on AM is given by the standard error of the mean, i.e., the standard deviation divided by the square root of the sample size. Fig.~\ref{fig:A23}, Fig.~\ref{fig:A45}, Fig.~\ref{fig:C23}, Fig.~\ref{fig:C45}, Fig.~\ref{fig:B12}, and Fig.~\ref{fig:B45} show the morphology of plane-of-sky magnetic fields inferred from Planck polarization, and VGT-PCA with resolution $\simeq 1 ^\circ$.

	In Fig.~\ref{fig:A23}, we highlight the region of the Taurus molecular cloud. We see the magnetic fields derived from VGT-PCA is well aligned with the magnetic fields inferred from Planck polarization. Earlier \citet{survey} applied the Velocity Channel Gradients technique (VChGs) \citep{LY18a} to study the magnetic fields in the Taurus molecular cloud using $^{13}$CO spectroscopic data. They showed that the magnetic fields inferred from $^{13}$CO molecular data are also statistically similar to the magnetic fields inferred from Planck polarization. Since both VGT-PCA and VChGs are tracing local magnetic fields using velocity gradients, it, therefore, encourages the study of the magnetic field properties in atomic and molecular gas, respectively. 
	
	In Fig.\ref{fig:B12}, we find there is a region, in which the relative angle between VGT-PCA and Planck is approximately $45^\circ$, which is theoretically unexpected. One possible reason is the fitting uncertainty from the sub-block averaging method. As explained in Sec.~\ref{sec:results}, there exists uncertainty in fitting Gaussian distribution, and the most probable value of Gaussian distribution has its standard deviation $\sigma$. Those factors would contribute to the overall uncertainty of magnetic field calculation \citep{IGs}. Subsequently, it was found that the $\sigma$ of the distribution is correlated with the statistical mean magnetization of the sub-region \citep{2018ApJ...865...46L}. In this case, the deviation is seen in Fig.\ref{fig:B12} upper right corner indicates that the properties or dynamics of \ion{H}{1} gas in that region are different. For example, the supernova will change the dynamics of gas in its surroundings. In addition, we list the expectation value $\mu$ of the histogram of the relative alignment between VGT-PCA and Planck 353GHz dust polarization, with resolution $\simeq 1^\circ$ in Tab.~\ref{tab:galfa}. We see that $\mu$ gets its maximum deviation from 0$^\circ$ at high-latitude regions, i.e., B2, B3, B4. Also, $\mu$ shows discrepancy when getting close the Galactic plane, i.e., A4, C2, C3. We propose the existence of molecular clouds near the Galacitc plane is one possible reason, since molecular clouds contribute to the dust polarization but not the \ion{H}{1}.

	\newpage
	
	\section{Simulation results}
	
	\begin{figure*}
		\centering
		\includegraphics[width=.99\linewidth,height=0.45\linewidth]{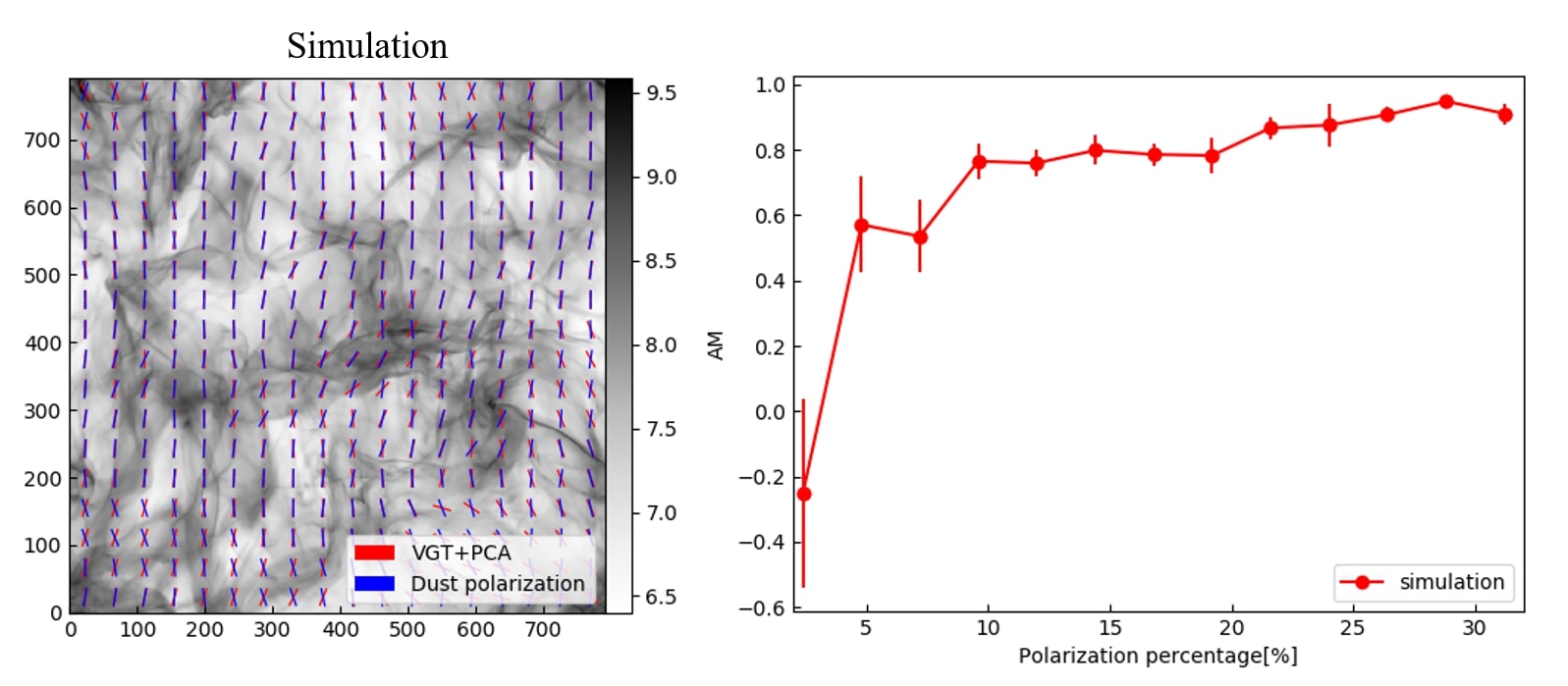}
		\caption{\label{fig:sim} The numerical testing results of predicting the dust polarization through the VGT+PCA technique. \textbf{Left}, the magnetic fields obtained from VGT+PCA (red segments) and dust polarization (blue segements), with overall AM=$0.78\pm0.03$. The MHD simulation imposes $M_S=6.14$ and Alfven Mach number $M_A=0.82$. \textbf{Right:} the variation of the AM with respect to the polarization percentage.}
	\end{figure*}
	
	We numerically test the ability of the VGT+PCA technique in predicting dust polarization. The numerical 3D MHD simulations are generated by ZEUS-MP/3D code \citep{2006ApJS..165..188H}, which uses a single fluid, operator-split, staggered grid MHD Eulerian assumption. To emulate a part of interstellar cloud, the periodic boundary conditions and solenoidal turbulence injections are applied in our simulations. We simulate the MHD turbulence using the barotropic equation of state, i.e., these clouds are isothermal with temperature T = 10.0 K, sound speed $c_s$ = 187 m/s, cloud size L=10 pc and initial density $\rho_0 \sim 884.23$ cm$^{-3}$. The Sonic Mach number $M_S$ is 6.14 and Alfven Mach number $M_A$ is 0.82 in our simulation.
	
	We follow the recipe used in Sec.~\ref{sec:method} to calculate the magnetic field morphology through either VGT+PCA or dust polarization. As shown in Fig.~\ref{fig:sim}, the numerical results also confirm that the robust ability of VGT+PCA in predicting dust polarization, with overall AM=$0.78\pm0.03$. We also plot the correlation between polarization percentage p defined in Eq.~\ref{eq.1} and the AM in corresponding regions. In Fig.~\ref{fig:sim}, we find the correlation coincides with the observational results, i.e., the AM is positively proportional to the polarization percentage when $p<10\%$. The AM gradually gets saturated when  
	$p>10\%$. It therefore numerically demonstrated our conclusion that insufficient polarization percentage or low dust grain alignment efficiency would cause the discrepancy between the magnetic fields obtained from gradients and dust polarization.

	
	
\end{document}